\documentclass[
aps,pra,showpacs,preprint
]{revtex4-1}

\usepackage{amsmath}
\usepackage{amssymb}
\usepackage{amsfonts}
\usepackage{graphicx}
\usepackage{epstopdf}

\begin{document} 

\title{A position-dependent mass harmonic oscillator \\ and deformed space}

\author{Bruno G. da Costa}
\email{bruno.costa@ifsertao-pe.edu.br} 
\affiliation{Instituto Federal de Educa\c c\~ao, Ci\^encia e Tecnologia do
             Sert\~ao Pernambucano, {\it Campus} Petrolina, 
             BR 407, km 08, 56314-520 Petrolina, Pernambuco, Brazil} 
\affiliation{Instituto de F\'isica, Universidade Federal da Bahia,
             Rua Barao de Jeremoabo, 40170-115 Salvador--BA, Brasil}
\author{Ernesto P. Borges} 
\email{ernesto@ufba.br} 
\affiliation{Instituto de F\'isica, Universidade Federal da Bahia,
             Rua Barao de Jeremoabo, 40170-115 Salvador--BA, Brasil}
\affiliation{National Institute of Science and Technology for Complex Systems, 
             Rua Xavier Sigaud 150, Rio de Janeiro 22290-180, Brazil}

\date{\today}

\begin{abstract}
We consider canonically conjugated
generalized space and linear momentum operators $\hat{x}_q$ and $ \hat{p}_q$ 
in quantum mechanics,
associated to a generalized translation operator 
which produces infinitesimal deformed displacements
controlled by a deformation parameter $q$.
A canonical transformation 
$(\hat{x}, \hat{p}) \rightarrow (\hat{x}_q, \hat{p}_q)$ 
leads the Hamiltonian of a position-dependent mass particle in usual space 
to another Hamiltonian of a particle with constant mass 
in a conservative force field of the deformed space.
The equation of motion for the classical phase space $(x, p)$ 
may be expressed in terms of the deformed (dual) $q$-derivative.
We revisit the problem of a $q$-deformed oscillator
in both classical and quantum formalisms.
Particularly, this canonical transformation 
leads a particle with position-dependent mass 
in a harmonic potential
to a particle with constant mass in a Morse potential.
The trajectories in phase spaces $(x,p)$ and $(x_q, p_q)$ are analyzed 
for different values of the deformation parameter.
Lastly, we compare the results of the problem in 
classical and quantum formalisms through 
the principle of correspondence and the WKB approximation.
\end{abstract} 

\pacs{03.65.Ca, 03.65.Ge, 03.65.Sq}


\maketitle

\section{Introduction}

During last three decades experimental and theoretical studies 
of quantum systems with position-dependent mass (PDM)
has attracted the interest of several researchers
of different areas. 
The most common example of an application of such systems  
is the motion of electrons and holes in semiconductors \cite{vonroos_1983}.
There are several others applications of quantum systems 
with PDM, for instance,
studies on inversion potential for NH$_{3}$ in density functional theory  
\cite{Aquino_1998},
He clusters \cite{Saavedra_1994},
optical properties of a spherical quantum dot \cite{Khordad_2012},
neutrino mass oscillations \cite{Bethe_1986}.

In many theoretical works related to systems with PDM,
the main goal is to obtain the eigenfunctions and energy levels 
by means of the Schr\"odinger equation for a system with 
a certain function of the mass in terms of the position 
and subject to a specific potential.
Generally, the Schr\"odinger equation 
for a system with PDM
is a non-homogeneous differential equation.
The most common approaches used to solve systems with PDM
are coordinate transformations and supersymmetry
\cite{plastino-1999,Amir-2016}.

Recently, Costa Filho {\it et al.} have proposed
a generalized translation operator which produces infinitesimal
nonlinear displacements, defined by
\cite{costa-filho-2011, costa-filho-2013, mazharimousavi,costa-borges}
\begin{equation}
\label{eq:translate}
	\hat{\mathcal{T}}_{q}
	(\varepsilon )|x\rangle 
  \equiv |{x + \varepsilon + \gamma_q x\varepsilon}\rangle.
\end{equation}
It may be associated with the $q$-algebra 
derived from the nonextensive formalism through 
$\gamma_q\equiv(1-q)/\xi$ \cite{lemans,borges_2004}.
The index $q$ is a dimensionless parameter, 
and $\xi$ is a characteristic length.
The operator $\hat{\mathcal{T}}_{q} (\varepsilon)$ leads to a generator 
of spatial translations corresponding 
to a position-dependent linear momentum given by 
\begin{equation}
 \label{eq:operator-momentum-generalized}
 \hat{p}'_q = (\hat{1}+\gamma_q \hat{x}) \hat{p},
\end{equation}
and consequently a particle with PDM.
This deformed momentum operator has been used to discuss a system 
with PDM for different potentials
in the quantum formalism
\cite{
costa-filho-2011, costa-filho-2013, 
mazharimousavi,costa-borges,
RegoMonteiro-Nobre-2013-PRA,
barbagiovanni-costafilho-2013,barbagiovanni-2014,
Vubangsi-Tchoffo-Fai-2014a,Vubangsi-Tchoffo-Fai-2014b,
Tchoffo-Vubangsi-Fai-2014c,Arda-Server}.
In these works, the deformed linear momentum operator is not Hermitian at
the coordinate basis $ \{|x \rangle \} $.
A modification can be done in order to make it Hermitian
\cite{mazharimousavi,
      costa-borges,
      Vubangsi-Tchoffo-Fai-2014a,
      Vubangsi-Tchoffo-Fai-2014b}: 
\begin{equation}
\label{eq:hermitian-operator-momentum-generalized}
\hat{p}_q = \frac{(\hat{1}+\gamma_q \hat{x}) \hat{p}}{2} +
   \frac{\hat{p}(\hat{1}+\gamma_q \hat{x})}{2}
         = (\hat{1}+\gamma_q \hat{x}) \hat{p} 
			- \frac{i\hbar \gamma_q}{2},
\end{equation}
with $[ \hat{x}, \hat{p} ] = i\hbar \hat{1}$.
Hermiticity leads to important properties: 
classical analogues for operators of dynamic variables,
real eigenvalues, and 
the 
existence of orthonormal basis of eigenstates.
This latter is related to the conservation 
law for probability in quantum mechanics.

Nonlinear generalizations of equations of motion 
formulated by means of the $q$-algebra
have been recently proposed
\cite{Nobre-RegoMonteiro-Tsallis-2011,
      Nobre-RegoMonteiro-Tsallis-2012,
      Plastino-Tsallis,
      RegoMonteiro-Nobre-2013-JMP,
      Plastino-Souza-Nobre-Tsallis-2014,
      Pennini-2014,
      Bountis-Nobre,
      Zamora-Rocca-Plasino-Ferri}.
Alternative generalizations of dynamic equations 
with Jackson's $q$-calculus can be found in the literature 
\cite{Macfarlane,
      Biedenharn,
      Lavagno,
      Lavagno-Gervino,Lavagno-2009}.
A classical deformed oscillator described by deformed
trigonometric functions has been introduced in \cite{borges_trig_1998}. 
Ref.\ \cite{costa-filho-2013} analyzed a quantum harmonic oscillator 
in a nonlinear space given by Eq.~(\ref{eq:translate}),
with the non-Hermitian operator (\ref{eq:operator-momentum-generalized}).
Deformed harmonic oscillators have nonlinear spectrum
and can describe diatomic molecules.
One important drawback of these formulations 
is the violation of the uncertainty principle
for certain values of the controlling parameter.

We  revisit the problem 
of a particle in a nonlinear space 
as described by Eq.~(\ref{eq:translate}),
and its linear momentum given by 
Eq.~(\ref{eq:hermitian-operator-momentum-generalized})
under the influence of the quadratic potential 
$V(x) = k x^2 / 2$, 
within both classical and quantum formalisms.
The paper is organized as follows:
we briefly review some properties of the $q$-algebra.
Next, we discuss the dynamics of a system in the nonlinear space 
for both classical and quantum formalisms.
Then, we solve the problem of a particle with PDM
in classical and quantum formalisms, 
comparing the results through the  WKB approximation
and the principle of correspondence.
Uncertainty principle is also analyzed.

\section{\label{sec:q-algebra}Deformed functions and derivatives}

The $q$-exponential is a generalization
of the ordinary exponential function,
defined by:
\begin{equation}
\exp_q{x} \equiv [1 + (1-q)x]_+^{1/(1-q)},
\end{equation} 
with $[A]_+=\max\{A,0\}$,
and the ordinary exponential is recovered as $q \rightarrow 1$
\cite{ct-quimicanova,Entropia-Tsallis,Tsallis-Springer-2009}.
It satisfies 
$  \exp_q (a) \exp_q (b) = \exp_q (a \oplus_q b) $
and
$ \exp_q (a) / \exp_q (b) = \exp_q (a \ominus_q b),$
where the symbol $\oplus_q$ represents the $q$-addition operator 
defined by ${a \oplus_q b} \equiv a + b + (1 - q) ab$, and
$\ominus_q$ represents the $q$-subtraction, 
${a \ominus_q b} \equiv \frac{a - b}{1 + (1 - q)b}$ \; ($b\ne (q-1)^{-1}$)
\cite{lemans,borges_2004}.
The inverse function of the $q$-exponential is the $q$-logarithm function, 
given by
\begin{equation}
 \label{eq:q-logatithm}
 \displaystyle \ln_q{x} \equiv \frac{x^{1-q} -1}{1-q}
 \qquad (x > 0).
\end{equation}

It is possible to define a generalization of the derivative
operator, based on these deformed algebraic operators
\cite{borges_2004}. Particularly,
\begin{eqnarray}
\label{eq:q-derivative}
\begin{array}{lll}
  \displaystyle D_{q} f(u) 
        &\equiv& \displaystyle 
		  \lim_{u'\to u}\frac{f(u') - f(u)}{u'\ominus_q u} 
\\
		& = & \displaystyle 
			\lim_{\Delta u \to 0}
			\frac{f(u \oplus_q \Delta u) - f(u)}{\Delta u} 
\\
        &=&      \displaystyle [1+(1-q)u] \frac{df(u)}{du},
 \end{array}
\end{eqnarray}
which the $q$-exponential is eigenfunction.
There is a dual $q$-derivative,
\begin{eqnarray}
\label{eq:q-derivative-dual}
\begin{array}{lll}
  \displaystyle \widetilde{D}_{q} f(u) 
        &\equiv& \displaystyle \lim_{u'\to u}\frac{f(u')\ominus_q f(u)}{u'- u} 
\\
        &=&      \displaystyle \frac{1}{1+(1-q)f(u)} \frac{df(u)}{du},
 \end{array}
\end{eqnarray}
that satisfies $\widetilde{D}_{q} \ln_q u = 1/u$.
These operators obey 
$\widetilde{D}_q x(y) = [D_q y(x)]^{-1}$,
$\forall q \in \mathbb{R}$.
The ordinary derivative is recovered for $q=1$ in both cases.

The deformed derivative operator $D_q f(u)$ may be
understood as the rate of variation of the function $f(u)$ with respect to 
a nonlinear variation of the independent variable 
$u \rightarrow u' =  
u \oplus_q \Delta u = u + \Delta u + (1 - q)u \Delta u$.
On the other hand, the dual deformed derivative operator $\widetilde{D}_q f(u)$
is the rate of a nonlinear variation of the function $f(u)$
with respect to the ordinary variation of the independent variable $u$.

Considering a real variable $u$, we have
\begin{equation}
d_q u \equiv \displaystyle \lim_{u'\to u} u' \ominus_q u 
      = \frac{du}{1 + (1 - q) u}.
\end{equation} 
The definition of the deformed variable $u_q$ (a deformed $q$-number)
\begin{equation}
\label{eq:transformation-u_q-u}
u_q \equiv \ln(\exp_q{u}) = \frac{\ln[1 + (1-q)u]}{(1-q)} 
\end{equation}
implies $d_q u = d u_q$, i.e.,
the \textit{deformed differential} of an ordinary variable $u$ 
may be expressed as
the ordinary differential of a \textit{deformed variable} $u_q$.
This $q$-deformed variable, Eq.~(\ref{eq:transformation-u_q-u}),
has already been defined in \cite{borges_trig_1998}.
It is curious to notice that Tsallis an R\'enyi entropies 
are related exactly through this transformation
(see Eq.~(8) of \cite{Entropia-Tsallis}).

The $q$-derivatives obey the following relations:
$D_{q} f(u) = df(u)/d_q u = df(u)/du_q$ and 
$\widetilde{D}_{q} f(u) = d_{q}f(u)/du = df_{q}(u)/du$,
{\it i.e.}, the deformed derivative of an ordinary variable
is equal to the ordinary derivative of the corresponding 
deformed variable.
The second $q$-derivatives must be used as
\begin{equation}
  D_{q}^2 f(u) 
        = [1+(1-q)u ] 
          \frac{d}{du}
             \left\{
                   [1+(1-q)u ] \frac{df}{du}
             \right\},
\end{equation}
and
\begin{equation}
  \widetilde{D}_{q}^2 f(u) 
        = \frac{1}{1+(1-q)f(u)} 
          \frac{d}{du}
             \left[
                   \frac{1}{1+(1-q)f(u)} \frac{df}{du}
             \right].
\end{equation}
Higher order deformed derivatives are evaluated accordingly.

\section{Dynamics of a system with position-dependent mass}

The deformed linear momentum operator (\ref{eq:operator-momentum-generalized})
is a generator of nonlinear translations. 
This operator is related to the $q$-derivative (\ref{eq:q-derivative}).
The effect of operator $\hat{p}'_q$ at coordinate basis 
$\{ | x \rangle \}$ on a state $| \alpha \rangle$ is
\begin{equation}
 \langle x|\hat{p}'_q |\alpha \rangle 
 = -i\hbar (1+\gamma_q x) \frac{d{\psi}(x)}{dx}  
 = -i \hbar D_{\gamma_q} {\psi}(x),
\end{equation}
where $ \psi(x) = \langle x | \alpha \rangle $
and $D_{\gamma_{q}} \equiv (1 + \gamma_{q}x)d/dx$.
Similarly, 
\begin{eqnarray}
 \langle x|\hat{p}_q |\alpha \rangle 
 & = & -i \hbar (1+\gamma_q x) \frac{d{\psi}(x)}{dx} 
       -\frac{i\hbar {\gamma}_{q}}{2} {\psi}(x)  
 \nonumber \\
 & = & -i \hbar D_{\gamma_q} {\psi}(x) 
       - \frac{i\hbar {\gamma}_{q}}{2} {\psi}(x).
\end{eqnarray}

Using the property
$ [\hat{p}, f(\hat{x})] = -i \hbar df (\hat{x}) / d\hat{x}$
with $ f(\hat{x}) = (\hat{1} + \gamma_q \hat{x})^{1/2} $, 
we can also write
$
 	\hat{p}_q = {(\hat{1} + \gamma_q \hat{x})^{1/2}
				\hat{p}
				(\hat{1} + \gamma_q \hat{x})^{1/2}}.
$
Thus,
\begin{equation}
 	\langle x|\hat{p}_q |\alpha \rangle 
	= \sqrt{1 + {\gamma}_q x} \left( \frac{\hbar}{i} \frac{d}{dx}\right) 
	\left[ \sqrt{1 + {\gamma}_q x} \psi (x) \right].
\end{equation}

Introducing a generalized space operator $\hat{x}_q $ 
canonically conjugated to $\hat{p}_q$, {\it i.e.}
$ 
 [\hat{x}_q, \hat{p}_q] = i\hbar \hat{1},
$
we get 
\begin{equation}
\label{eq:operator-position-generalized}
 \hat{x}_q = \frac{\ln (\hat{1}+\gamma_q \hat{x})}{\gamma_q} 
                = \xi \ln[\exp_q ( \hat{x}/\xi) ].
\end{equation}
In particular, 
$(\hat{x}, \hat{p}) \longrightarrow (\hat{x}_q, \hat{p}_q)$
forms a canonical transformation that leads 
a $q$-addition of two positions at basis $\{ | x \rangle \}$ 
into a usual addition of two positions in a deformed space at basis 
$\{ | x_q \rangle \}$, 
{\it i.e.}, $ x' \oplus_q x \longrightarrow x'_q + x_q $.
These hermitian operators present the following classical analogs:
\begin{eqnarray}
\label{eq:canonical-transf}
    \left\{ 
   \begin{array}{ll} 
    \displaystyle p_q = (1 + \gamma_q x)p \\
    \\
    \displaystyle x_q = \frac{\ln (1+\gamma_q x)}{\gamma_q} =
    \xi \ln \left[ \exp_{q} (x/\xi) \right] \\ 
    \end{array}
    \right.
\end{eqnarray}
with the generating function given by
$\phi(x_q, p) = - p(e^{\gamma_q x_q} + 1)/ \gamma_q.$
In the following, we analyze some implications 
of this canonical transformation
in both classical and quantum formalisms.

\subsection{ Deformed classical formalism}

Let us initially address the classical problem of a constant mass particle 
submitted to a conservative force with potential $V(x_q)$,
and the linear deformed momentum $p_q$, whose Hamiltonian is 
\begin{equation}
\label{eq:hamiltonian_xgamma_pgamma}
 K(x_q , p_q ) = \frac{p_q ^2}{2 m_0} +  V(x_q).
\end{equation}

The canonical transformation given by Eqs.~(\ref{eq:canonical-transf})
leads to the new Hamiltonian (see, for instance, \cite{cruz_y_cruz_2013})
\begin{equation}
\label{eq:hamiltonian_x_p}
 H(x , p) = \frac{p^2}{2m(x)} +  V(x),
\end{equation}
with
%
\begin{equation}
\label{eq:m(x)}
 m(x) = \frac{m_0}{(1+\gamma_q x)^2}.
\end{equation}
The equation of motion is 
\begin{equation}
 \label{eq:dp/dt}
 \dot{p} = - \frac{\gamma_q (1+\gamma_q x)p^2}{m_0} - \frac{dV(x)}{dx},
\end{equation}
with
$p = m(x) \dot{x}$,
thus
\begin{eqnarray} \label{eq:equation-of-motion} 
m_0 \left[ \frac{\ddot{x}}{(1+\gamma_q x)^2} - 
    \frac{\gamma_q \dot{x}^2}{(1+\gamma_q x)^3}  \right]  = - \frac{dV(x)}{dx}.
\end{eqnarray}

This equation may be conveniently rewritten as
\begin{equation}
\label{eq:second_newton_law_generalized}
       m_0 \widetilde{D}^2_{\gamma_q}   x (t) = F(x),
\end{equation}
{\it i.e.}, a deformed Newton's law for a space with nonlinear displacements.
The generalized displacement of a PDM $m(x)$
in a usual space ($d_q x$) is mapped into a constant mass $m_0$
in a deformed space with usual displacement ($d x_q$):
$\displaystyle d_q x \equiv
    {\xi} \left[ \left( \frac{x+dx}{\xi} \right) 
    \ominus_q 
    \left( \frac{x}{\xi} \right) \right] 
    = \frac{dx}{1+\gamma_q x} \equiv d x_q$. 
The time evolution is, thus, governed by the generalized dual derivative,
$\displaystyle \widetilde{D}_{\gamma_q} x = \frac{1}{1+\gamma_q x}\frac{dx}{dt}$.

The particle velocity,
\begin{equation}
\label{eq:velocity-energy}
	\dot{x} 
	= \sqrt{\frac{2}{m(x)} [E - V(x)]},
\end{equation}
may be rewritten as a deformed particle velocity:
\begin{equation}
\label{eq:deformed-velocity-energy}
	\widetilde{D}_{{\gamma}_q} x =
	\sqrt{\frac{2}{m_0} [E - V(x)]}.
\end{equation}
Coherently, the particle position can be obtained through the $q$-integral
\begin{eqnarray}
\label{eq:integral-x(t)}
	t - t_0
	& = &
	\displaystyle{
	\pm \int_{x_0}^{x} \frac{dx}{\sqrt{\frac{2}{m(x)}[E - V(x)]}} 
	} \nonumber \\
	& = &
	\displaystyle{
	\pm \int_{x_0}^{x} \frac{dx}{(1 + {\gamma}_q x)\sqrt{\frac{2}{m_0}[E - V(x)]}}
	} \nonumber \\
	& = &
	\displaystyle{\pm \int_{x_0}^{x} \frac{d_q x}{\sqrt{\frac{2}{m_0}[E - V(x)]}}}.
\end{eqnarray}
%

\subsection{Deformed quantum formalism}

Consider a quantum system described by the Hamiltonian 
$ \hat{K}(\hat{x}_q,\hat{p}_q) = \hat{p}_q^2/2m_0 + V(\hat{x}_q) $,
at the coordinate basis $\{|x_q \rangle\} $.
For the ket state $|\alpha (t) \rangle$, 
the Schr\"odinger equation is
$
i\hbar \frac{\partial}{\partial t} | \alpha (t) \rangle
 = \hat{K} |\alpha (t) \rangle,
$
{\it i.e.}
\begin{equation}
\label{eq:equation-of-schroedinger-basis-x-q-wave-function}
	 i\hbar \frac{\partial \Phi (x_q, t)}{\partial t} = 
	-\displaystyle \frac{{\hbar}^2}{2m_0} 
	\frac{\partial^{2} \Phi (x_q, t)}{\partial x_q^2}
	+ V(x_q)\Phi (x_q, t)
\end{equation}
with $\Phi(x_q, t) \equiv \langle x_q |\alpha (t) \rangle $.

From the canonical transformation given by
Eqs.~(\ref{eq:hermitian-operator-momentum-generalized})
and (\ref{eq:operator-position-generalized}), 
the Hamiltonian operator at basis $\{|x \rangle\}$ is
\begin{eqnarray}
 \label{eq:eq-of-schroedinger-basis-x}
	\hat{H}(\hat{x}, \hat{p}) & = &
	\displaystyle
	\frac{1}{2m_0}
	\Biggl[ 
	\frac{(\hat{1}+\gamma_q \hat{x}) \hat{p}}{2} +
	\frac{\hat{p}(\hat{1}+\gamma_q \hat{x}) }{2} 
	\Biggr]^2 
    + V(\hat{x}) 
	\nonumber \\
	&=&
	\displaystyle
	\frac{1}{2m_0}
	\left[(\hat{1}+\gamma_q \hat{x})^{1/2} 
	\hat{p}(\hat{1}+\gamma_q \hat{x})
	\hat{p}(\hat{1}+\gamma_q \hat{x})^{1/2} \right]
	+ V(\hat{x}).
\end{eqnarray}
It is in agreement with kinetic operator introduced by
von Roos \cite{vonroos_1983}
for systems with a PDM operator given by
$m(\hat{x}) = m_0/(\hat{1} + \gamma_q \hat{x})^2$,
and it can be rewritten as
\begin{equation}   
\label{eq:general-hamiltonian-pdm}
	\hat{H}(\hat{x}, \hat{p}) = 
	-\frac{\hbar^2}{2} 
	\biggl\{
	[m(\hat{x})]^{\varsigma} \frac{d}{dx} [m(\hat{x})]^{\zeta} \frac{d}{dx} [m(\hat{x})]^{\varsigma} 
	\biggr\}
	+ V(\hat{x}),
\end{equation}
with $\varsigma = \frac{\zeta}{2} = -\frac{1}{4}$.
The particular case $(\varsigma, \zeta) = (-\frac{1}{4}, -\frac{1}{2})$ 
maps the potential $V(\hat{x})$ into an effective potential $V(\hat{x}_q)$, 
that is independent on $m(\hat{x})$ \cite{cruz_y_cruz_2009}.

The Schr\"odinger equation
$
i\hbar \frac{\partial}{\partial t} | \alpha (t) \rangle
= \hat{H} | \alpha (t) \rangle 
$
may be explicitly written in terms of the wave function 
$\Psi(x, t) \equiv \langle x | \alpha  (t) \rangle$
as
\begin{equation}
\label{eq:schroedinger-equation-sho-pdm2}
i\hbar \frac{\partial \Psi (x,t)}{\partial t} =
-\frac{{\hbar}^2 ( 1+\gamma_q x)^2}{2m_0} \frac{\partial^2 \Psi (x,t)}{\partial x^2}
-\frac{{\hbar}^2\gamma_q (1+\gamma_q x)}{m_0}\frac{\partial \Psi (x,t)}{\partial x} 
-\frac{{\hbar}^2\gamma_{q}^2}{8m_0}\Psi (x,t)  
+ V(x) \Psi (x,t).
\end{equation}

The probability density $\rho(x,t) \equiv  |\Psi(x,t)|^2$
obeys the continuity equation
\begin{equation}
\label{eq:contituity-equation-pdm-sytem}
	\frac{\partial \rho (x,t)}{\partial t} + 
	\frac{\partial J(x,t)}{\partial x} = 0,
\end{equation}
where the current density is given by  
\begin{equation}
\label{eq:current-density}
	J(x,t) \equiv 
		\operatorname{Re}
		\left\{
		\Psi^{\ast} (x,t) \left( \frac{\hbar}{i}
		\frac{\partial}{\partial x} \right)
		\left[\frac{1}{m(x)} \Psi (x,t)\right]
		\right\}.
\end{equation}
Eq.~(\ref{eq:schroedinger-equation-sho-pdm2}) 
may be conveniently rewritten by means of the transformation
commonly use for PDM systems,
$ \Psi (x, t) = {[m(x)/m_0]}^{1/4} 
\Phi_q (x, t)$ 
\cite{cruz_y_cruz_2009},  
that results
\begin{equation}
\label{eq:transf-psi-varphi}
	\displaystyle \Psi (x, t) \equiv 
			\frac{
		\Phi_q (x,t)
				}{\sqrt{1 + \gamma_q x}},
\end{equation}
and
\begin{equation}
\label{eq:schroedinger-equation-sho-pdm}
	i\hbar \frac{\partial \Phi_q (x, t)}{\partial t} =
	-\frac{\hbar^2 (1 + \gamma_q x)^2}{2m_0} 
	\frac{\partial^{2}\Phi_q (x, t)}{\partial x^2}
	-\frac{\hbar^2 \gamma_q (1 + \gamma_q x)}{2m_0}
	\frac{\partial \Phi_q (x, t)}{\partial x} 
	+ V(x) \Phi_q (x, t),
\end{equation}
or, more compactly,
\begin{equation}
\label{eq:deformed-schroedinger-equation}
	i\hbar \frac{\partial \Phi_q (x, t)}{\partial t} = 
	-\frac{\hbar^2}{2m_0} \mathcal{D}_{\gamma_q}^2 \Phi_q (x, t) 
	+ V(x) \Phi_q (x, t).
\end{equation}
This equation is equivalent to
Eq.~(\ref{eq:equation-of-schroedinger-basis-x-q-wave-function})
with 
$x \longrightarrow x_q = \xi \ln[\exp_q (x/\xi) ],$
$\mathcal{D}_{\gamma_q} \equiv \partial / \partial x_q
			= (1 + \gamma_q x) \partial / \partial x$,
and $\Phi_q (x, t) = \Phi (x_q(x), t)$.
It is also equivalent to 
Eq.~$(14)$ of \cite{costa-filho-2011} 
which corresponds to a Schr\"odinger-like equation for a field $\Phi_q (x,t)$
with the non-hermitian operator $\hat{p}'_q$,
and an associated non-Hermitian Hamiltonian operator given by
\begin{eqnarray}
\label{eq:operator-H'}
\hat{H'} &=& \frac{1}{2m_0} (\hat{p}'_q)^2 + V(\hat{x}) \nonumber \\
		 &=& \frac{1}{2m_0}(1 + \gamma_q \hat{x})\hat{p}(1 + \gamma_q \hat{x})\hat{p} 
			+ V(\hat{x}).
\end{eqnarray}
There is an equivalence between describing  
a Hermitian system with kinetic energy term 
of a PDM and a non-Hermitian one 
with kinetic energy term deformed 
in terms of the spatial $q$-derivative.
Furthermore, the  field  $\Psi(x, t)$ is replaced by a new deformed field 
$\Phi_q (x,t)$.
Similar to the classical formalism,
the Schr\"odinger-like equation (\ref{eq:deformed-schroedinger-equation})
for a system with PDM may also be written in terms of the $q$-derivative.

Note that if the field $\Phi(x_q, t)$ is the solution of the Schr\"odinger 
equation at basis $\{ | x_q \rangle \}$, 
then the field $\Psi (x, t)$ is its solution at basis 
$\{ | x \rangle \}$. 
Thus, the Schr\"odinger equation for the field $\Psi (x,t)$ 
for a system with PDM in a usual space 
$\{ | x \rangle \}$
is mapped into an equation for the field $\Phi (x_q,t)$ 
in a deformed space $ \{|x_q \rangle\} $.
Furthermore, 
if $\Phi (x_q, t)$ is normalized, then $\Psi (x, t)$ is also normalized.
In fact,
\begin{equation}
\label{eq:normalization}
	\int_{x_{q, i}}^{x_{q, f}} \Phi^{\ast} (x_q, t) \Phi (x_q, t) dx_q 
        =
	\int_{x_i}^{x_f} \frac{\Phi_q^{\ast} (x, t) \Phi_q (x, t)}
                              {1 + \gamma_q x} dx 
        = 1
\end{equation}
and using Eq.~(\ref{eq:transf-psi-varphi}), we have
$\int \Psi^{\ast} (x, t) \Psi (x, t) dx = 1.$
The deformed space implies a deformed metric, with the deformed inner
product defined by the $q$-integral 
(see \cite{costa-filho-2013}):
\begin{eqnarray}
\langle \varphi_{b} (x) | \varphi_{a} (x) \rangle_{q}
	& \equiv & \int_{x_i}^{x_f} \frac{\varphi_{b}^{\ast} (x) \varphi_{a}(x)}{1 + \gamma_q x} dx
	\nonumber \\ 
	& = & \int_{x_i}^{x_f} \varphi_{b}^{\ast} (x) \varphi_{a}(x) d_{q} x
 	\nonumber \\ 
 	& = &   \int_{x_{q, i}}^{x_{q, f}} \varphi_{b}^{\ast} (x_q) \varphi_{a}(x_q) d x_{q}
 	\nonumber \\ 
	& = & \langle \varphi_{b} (x_q) | \varphi_{a}(x_q) \rangle .  
\end{eqnarray}
The deformed continuity equation reads:
\begin{equation}
 \label{eq:deformed-contituity-equation}
  \frac{\partial \varrho_q(x,t)}{\partial t}
  + \mathcal{D}_{\gamma_q} \mathcal{J}_q(x, t) 
  = 0,
\end{equation}
with  
\begin{equation}
 \label{eq:deformed-current-density}
	\mathcal{J}_q(x, t) \equiv 
		\operatorname{Re}
		\left[
		\Phi_q^{\ast} (x, t) 
		\left( \frac{\hbar}{i}
		\mathcal{D}_{\gamma_q}\right)
		\left( \frac{\Phi_q (x, t)}{m_0} \right)
		\right].
\end{equation}
Consistently, 
Eqs.~(\ref{eq:contituity-equation-pdm-sytem})--(\ref{eq:transf-psi-varphi})
lead to 
(\ref{eq:deformed-contituity-equation})--(\ref{eq:deformed-current-density}).

The quantum formalism for PDM in terms of the field $\Phi_q(x,t)$
replaces the usual derivative and integral operators 
(with respect to the spatial variable $x$)
by the $q$-derivative and $q$-integral.
The same feature applies for the classical formalism, 
but the with the dual $q$-derivative, instead.
This is due to the fact that in the quantum formalism, 
the equations that describe the dynamics of the system 
(such as the Schr\"odinger equation), 
takes into account nonlinear spatial variations 
of the independent variable $x$ ($\Phi=\Phi_q (x, t)$),
which is directly related to the $q$-derivative (\ref{eq:q-derivative}).
On the contrary, in the classical formalism, 
the nonlinear spatial variation takes place
on the dependent variable $x(t)$, 
directly associated with the definition of the dual $q$-derivative 
(\ref{eq:q-derivative-dual}).

According to Ehrenfest's theorem, the time evolution of the expectation values
of the space $\hat{x}$ and linear momentum $\hat{p}$ operators are 
respectively given by
\begin{subequations}
 \begin{equation}
  \label{eq:d<x>/dt}
  \frac{d \langle \hat{x} \rangle}{dt} = 
   \frac{\langle 
		  (\hat{1}+\gamma_q \hat{x})
 	      \, \hat{p} \,
 		  (\hat{1}+\gamma_q \hat{x}) 
		\rangle}
	   {m_0}, 
 \end{equation}
 and
 \begin{equation}
  \label{eq:d<p>/dt}
  \frac{d \langle \hat{p} \rangle}{dt} =
   - \frac{\gamma_q}{m_0}\langle 
	 \hat{p} \, (\hat{1}+\gamma_q \hat{x}) \, \hat{p} \rangle 
   - \Biggl\langle \frac{dV}{d\hat{x}} \Biggl\rangle,
 \end{equation}
\end{subequations}
with $\int J(x, t) dx = d\langle \hat{x} \rangle / dt$.

\section{Classical harmonic oscillator with position-dependent mass}

Consider a particle with mass given by Eq.~(\ref{eq:m(x)}) 
under the influence of a quadratic potential, whose the Hamiltonian is
\begin{equation}
\label{eq:hamiltonin-pdm-classical}
	H(x, p) = \frac{p^2}{2m(x)} + \frac{1}{2}kx^{2}.
\end{equation}

The deformed Newton's law for this problem is:
\begin{equation}
\label{eq:second_newton_law_generalized-linear-force}
      \widetilde{D}^2_{\gamma_q}   x (t) = - \omega_0^2 x,
\end{equation}
where the angular frequency $ \omega_0 = \sqrt{k/m_0} $ 
corresponds to the usual case $\gamma_q = 0$.

\begin{figure}[!htb]
 \centering
 \begin{minipage}[b]{0.48\linewidth}
  \includegraphics[width=\linewidth]{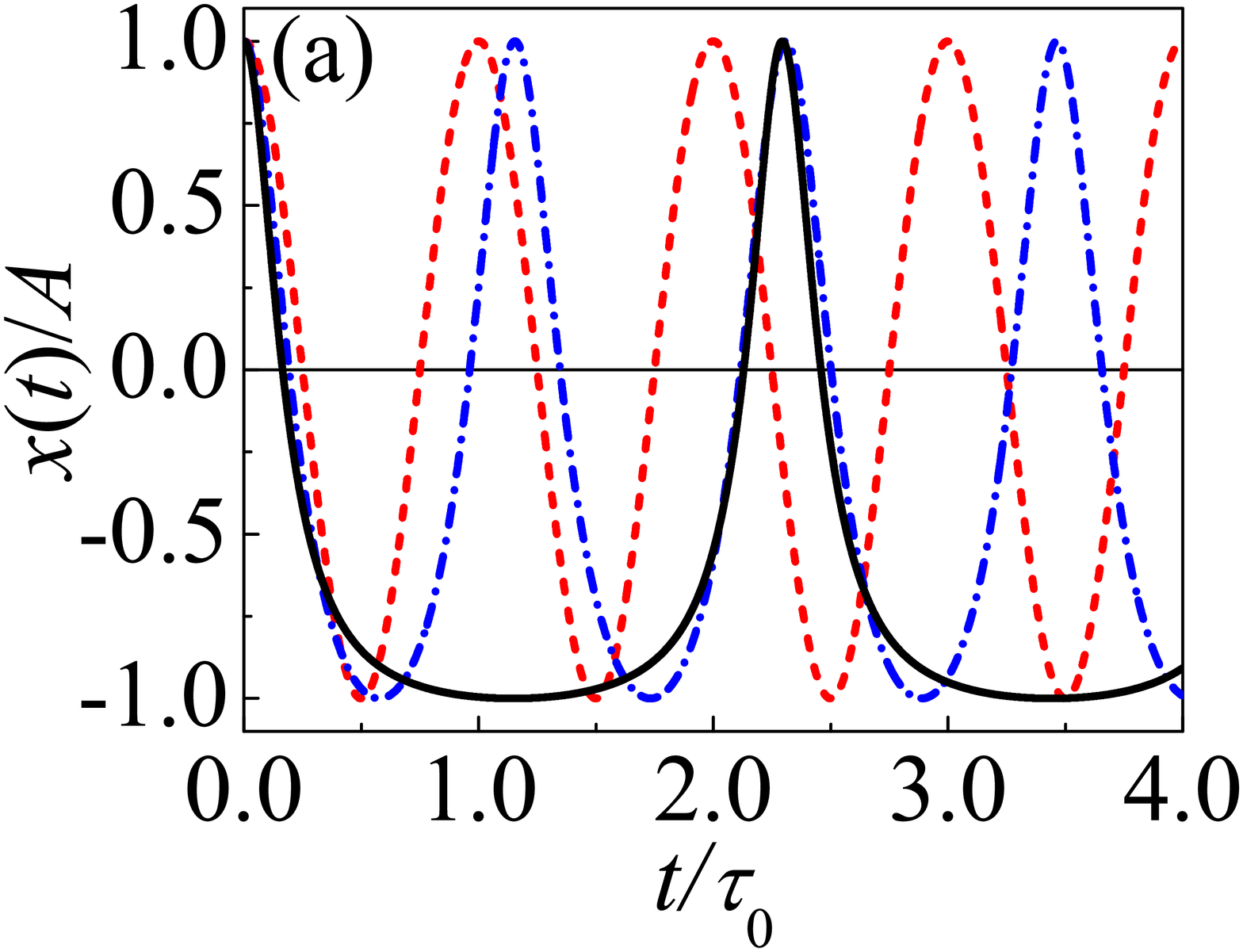}
 \end{minipage} 
 \begin{minipage}[b]{0.48\linewidth}
  \includegraphics[width=\linewidth]{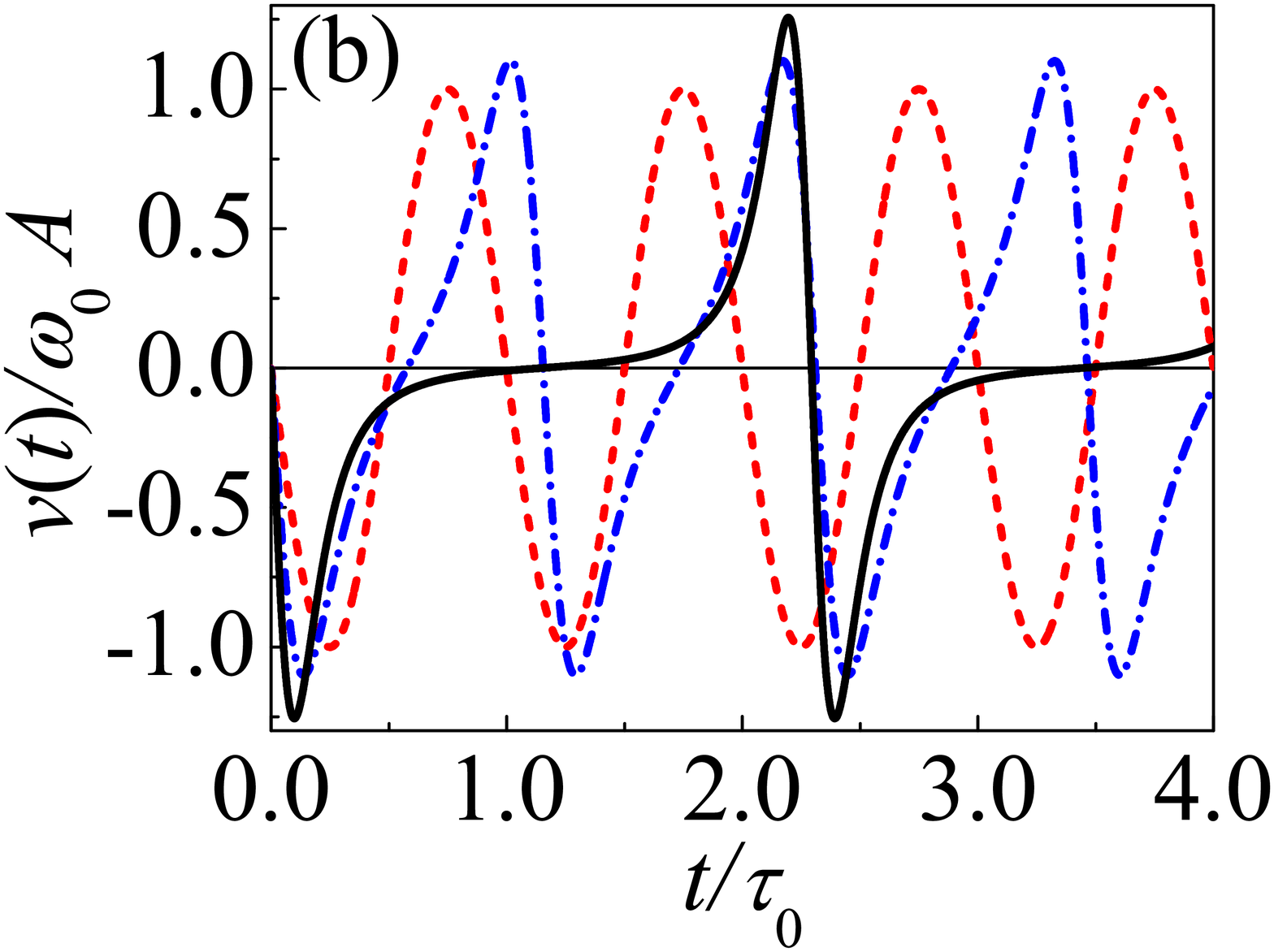}
 \end{minipage} \\
 \begin{minipage}[b]{0.48\linewidth}
  \includegraphics[width=\linewidth]{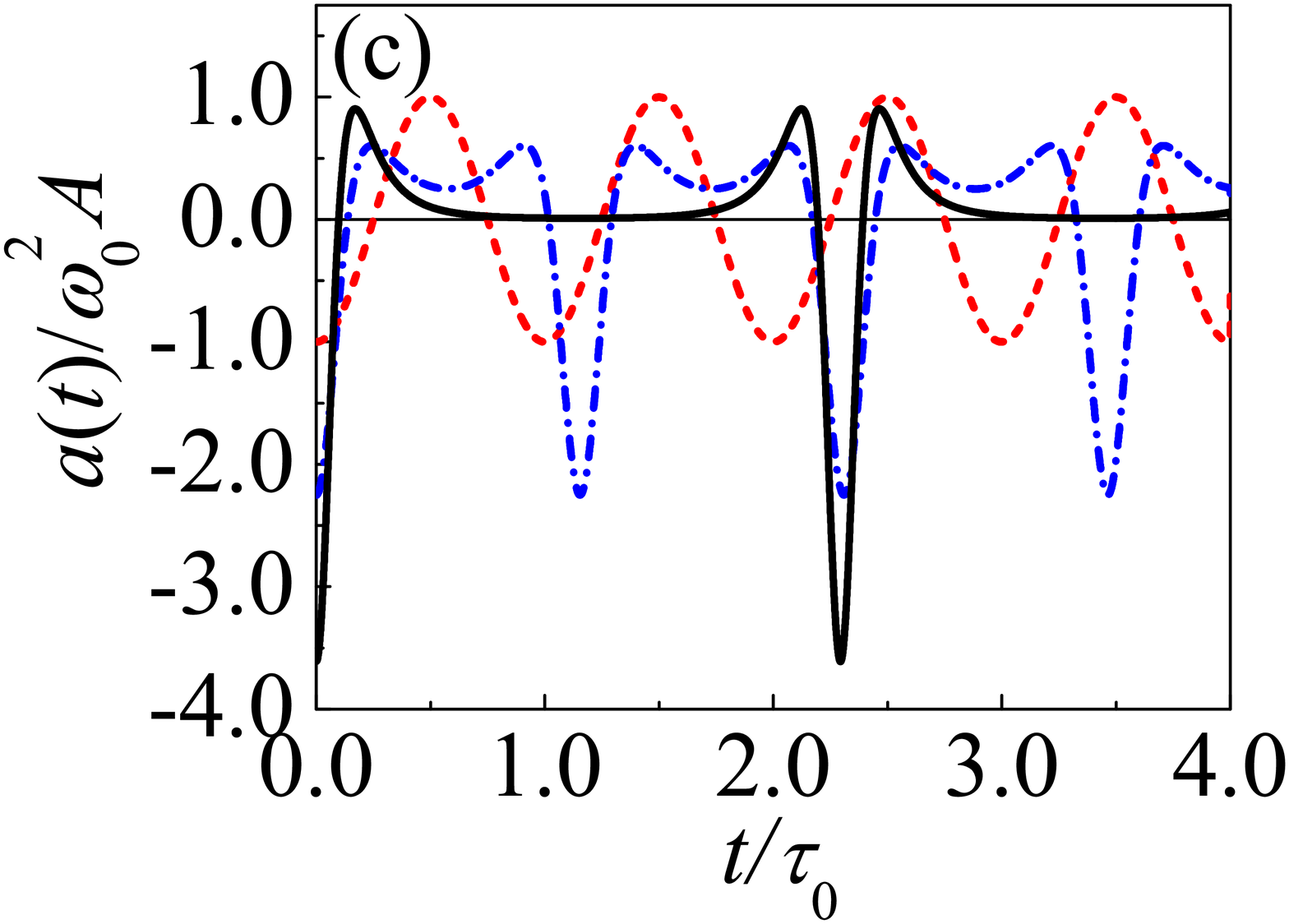}
 \end{minipage} 
 \begin{minipage}[b]{0.48\linewidth}
  \includegraphics[width=\linewidth]{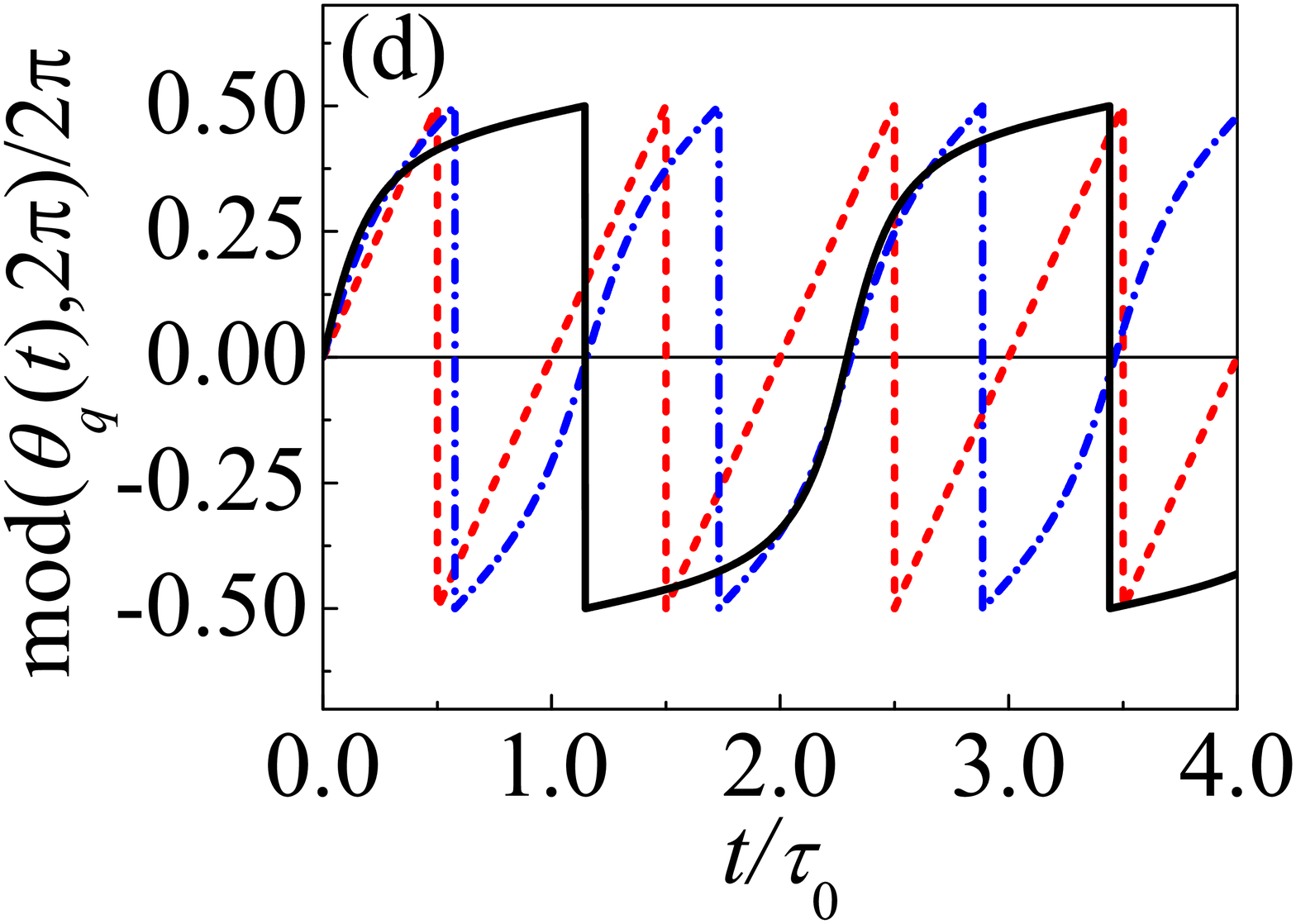}
 \end{minipage}
 \caption{\label{fig:xvaphi} 
  (Color online) Temporal evolution of 
  (a) position, 
  (b) velocity,
  (c) acceleration and 
  (d) phase 
  for a PDM classical oscillator
  given by Eq.~(\protect\ref{eq:m(x)}).
  $\gamma_q A = $ 0 (dashed red), 0.5 (dash-dotted blue), and 0.9 (solid black), 
  ($\tau_0=2\pi / \omega_0$, $\delta = 0$).
 }
\end{figure}

The velocity,
$v(t) = \dot{x} = \pm(1 + \gamma_q x) \omega_0 \sqrt{A^2 - x^2},$
is rewritten as
\begin{equation}
\label{eq:velocity_SHO_and_dual_derivate}
       \widetilde{D}_{\gamma_q} x (t) = \pm \omega_0 \sqrt{A^2 - x^2},
\end{equation}
where $A$ is the amplitude of oscillations.
The solution of 
Eq.~(\ref{eq:second_newton_law_generalized-linear-force}), or
Eq.~(\ref{eq:velocity_SHO_and_dual_derivate}), is 
\begin{equation}
\label{eq:equation-of-position}
         x(t)  = A \cos \left[ \theta_q(t) \right]
\end{equation}
with
$$
\theta_q(t) =
  2\ \textrm{atan}
  \left[ 
    \sqrt \frac{1 + \gamma_q A}{1 - \gamma_q A} 
    \;
    \textrm{tan}\left( 
                  \sqrt{1-\gamma_q ^2 A ^2} 
                  \ 
                  \frac{\omega_0 t + \delta}{2} 
                \right) 
  \right],
$$
that is a periodic function with period 
$\tau_q = \frac{2\pi}{\omega_0 \sqrt{1 - \gamma_q^2 A^2}}$.
Figure \ref{fig:xvaphi} shows position, velocity, acceleration and phase 
for $0 \leq \gamma_q A  < 1$ 
(the usual case $\gamma_q A = 0$ is illustrated for comparison).
As $\gamma_q A$ approaches $1$, the particle remains close to the position 
$x =-A$ for longer periods of time,
since the mass is increased in this region.

\begin{figure}[!htb]
 \centering
 \begin{minipage}[h]{0.48\linewidth}
  \includegraphics[width=\linewidth]{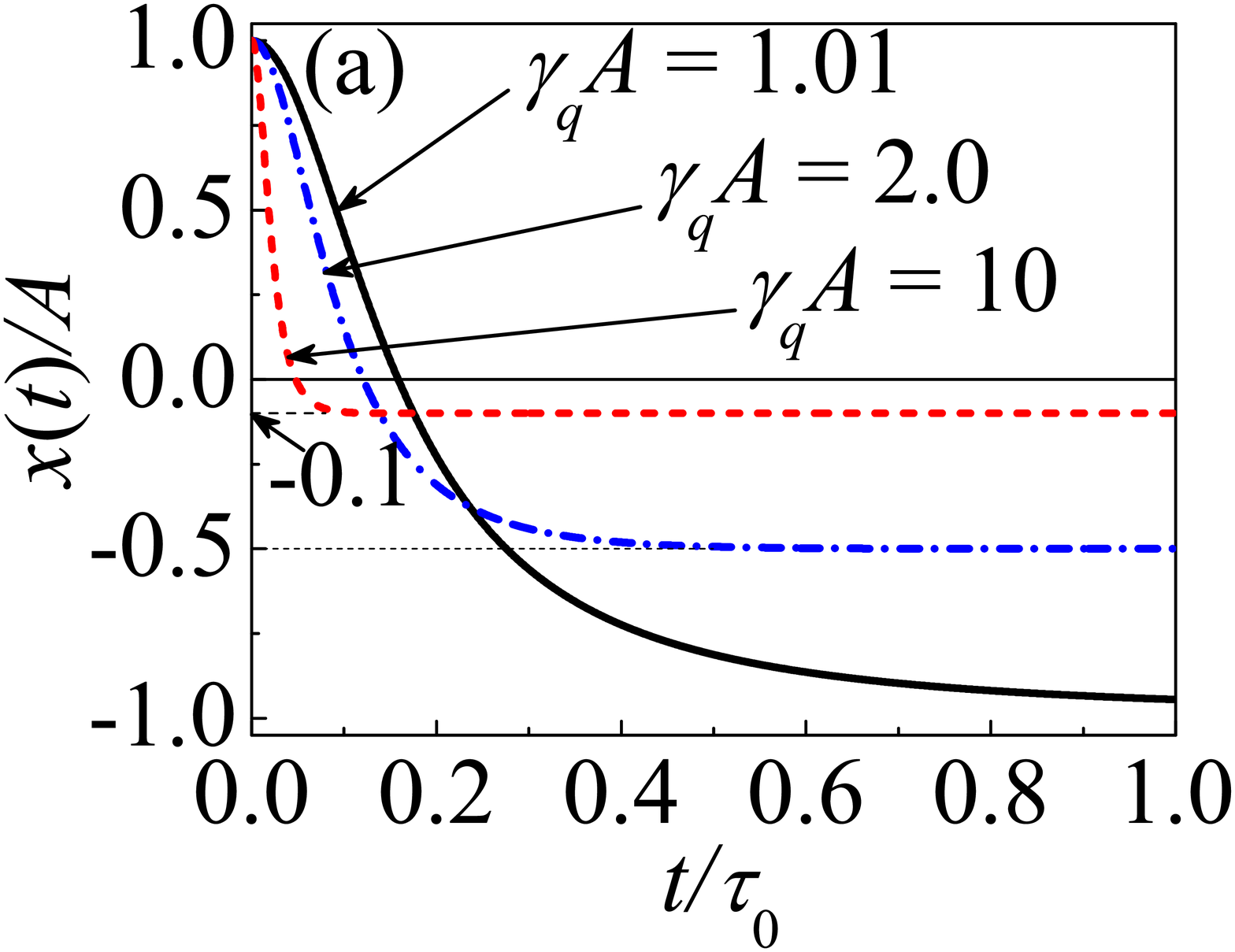}
 \end{minipage}
 \begin{minipage}[h]{0.48\linewidth}
  \includegraphics[width=\linewidth]{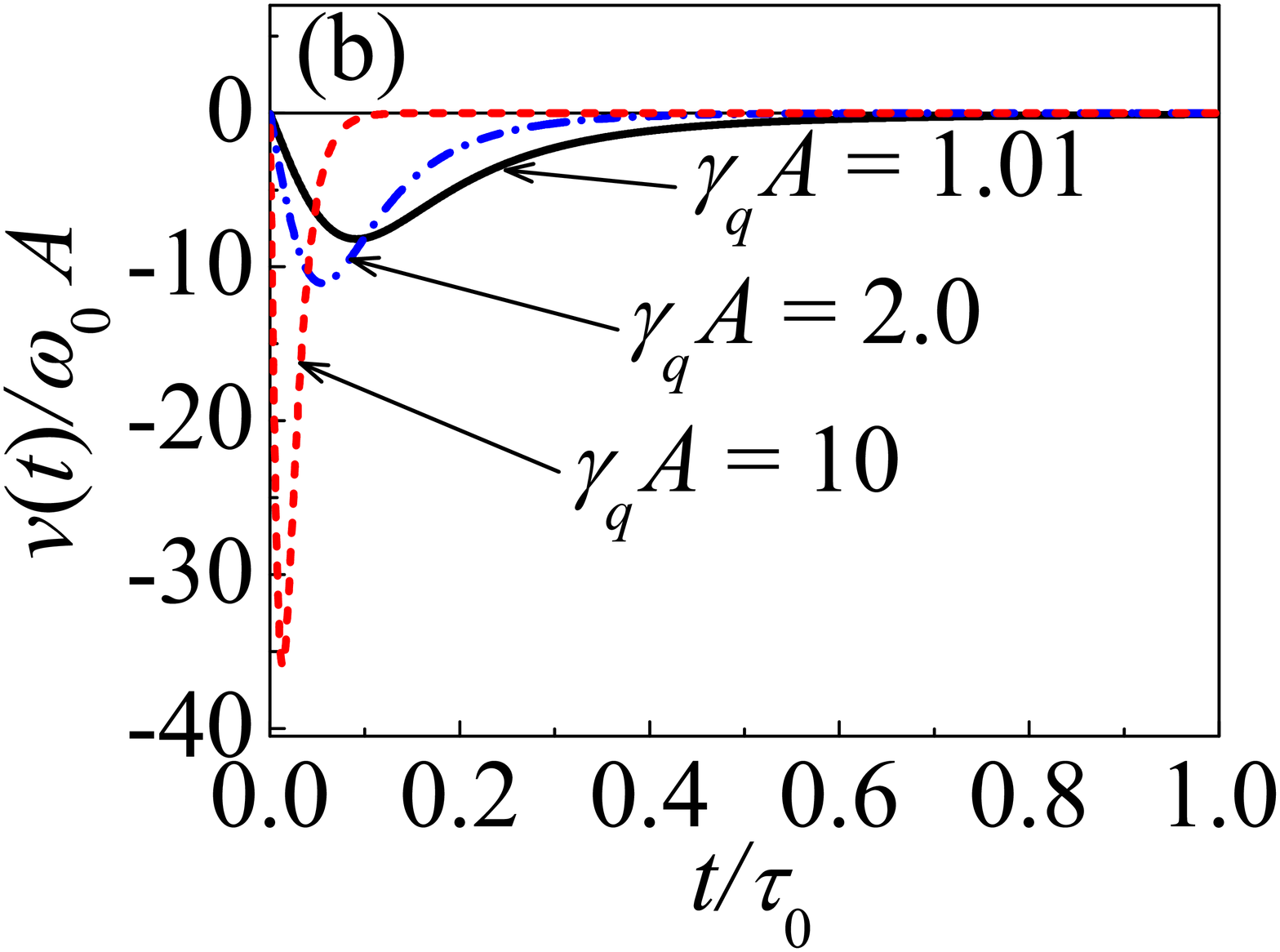}
 \end{minipage}
 \caption{\label{fig:gamma_q>1} 
  (Color online)
  Temporal evolution of position and velocity 
  for a PDM classical oscillator
  (Eq.~(\protect\ref{eq:m(x)})).
  $\gamma_qA =$ 1.01 (solid black), 2.0 (dash-dotted blue), and 10 (dashed red).
 }
\end{figure}

For $\gamma_q A > 1$, the system looses its oscillatory nature and the particle 
moves between $-1/\gamma_q < x \leq A$,
as illustrated in Figure \ref{fig:gamma_q>1}. 
When $t \gg  \tau_0$, $x$ asymptotically approaches 
$x_{\text{min}}=-1/\gamma_q $. 
Figure \ref{fig:gamma_q>1} also shows 
the particle velocity as a function of time.
For $t> t^\ast$, 
where
\begin{equation}
t^\ast = \frac{\tau_q}{\pi} \textrm{atanh} 
		\left[ 
		\frac{(\gamma_q A-1)(1+4\gamma_q A - \sqrt{1 + 8\gamma_q^2 A^2})}
		{(\gamma_q A+1)(1+4\gamma_q A + \sqrt{1 + 8\gamma_q^2 A^2})} 
		\right]^{1/2},
\end{equation}
the absolute value of the speed gradually decreases 
and eventually the particle comes to rest at $t \to \infty$.

\begin{figure}[!t]
 \centering
 \begin{minipage}[h]{0.49\linewidth}
  \includegraphics[width=\linewidth]{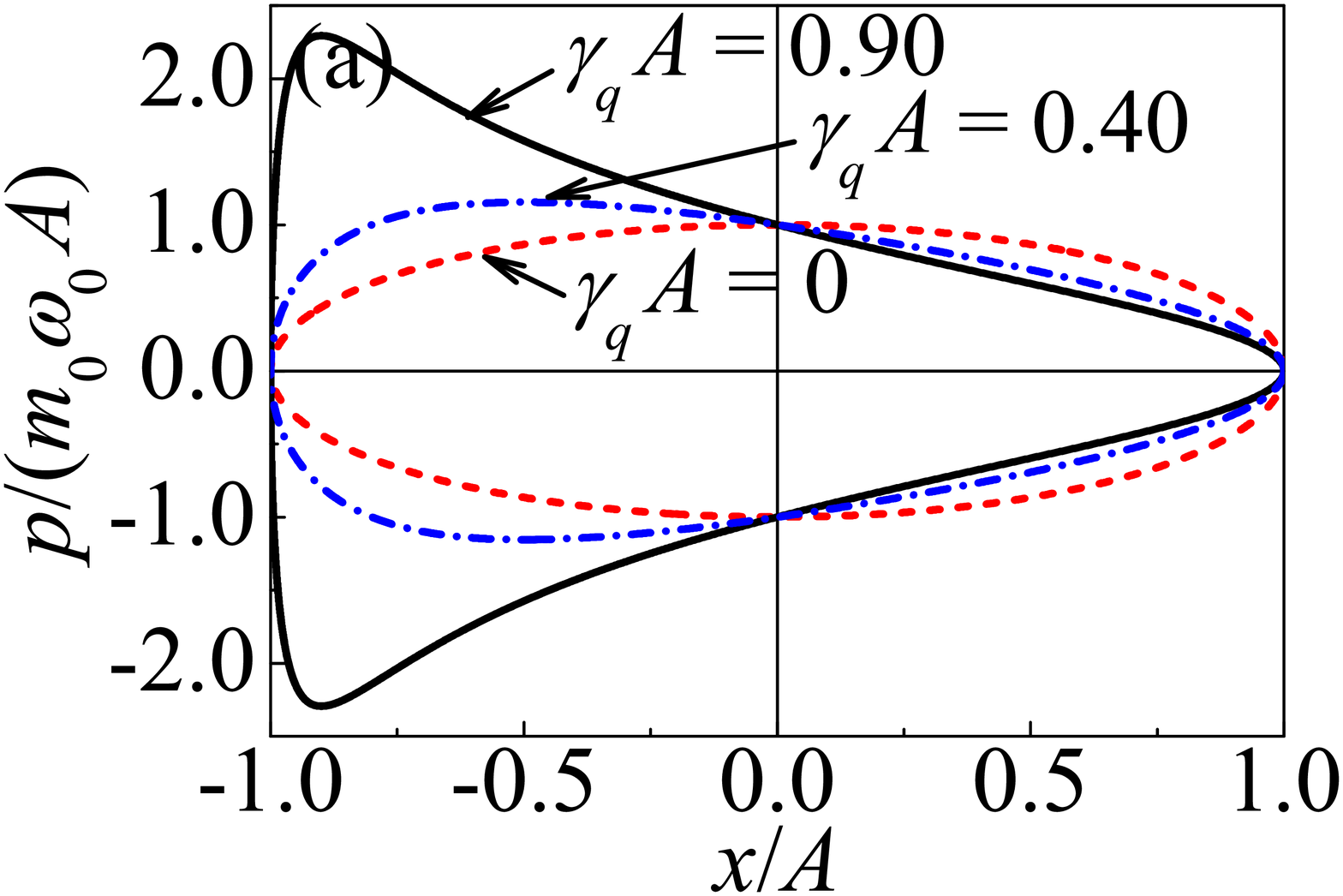}
 \end{minipage}
 \begin{minipage}[h]{0.49\linewidth}
  \includegraphics[width=\linewidth]{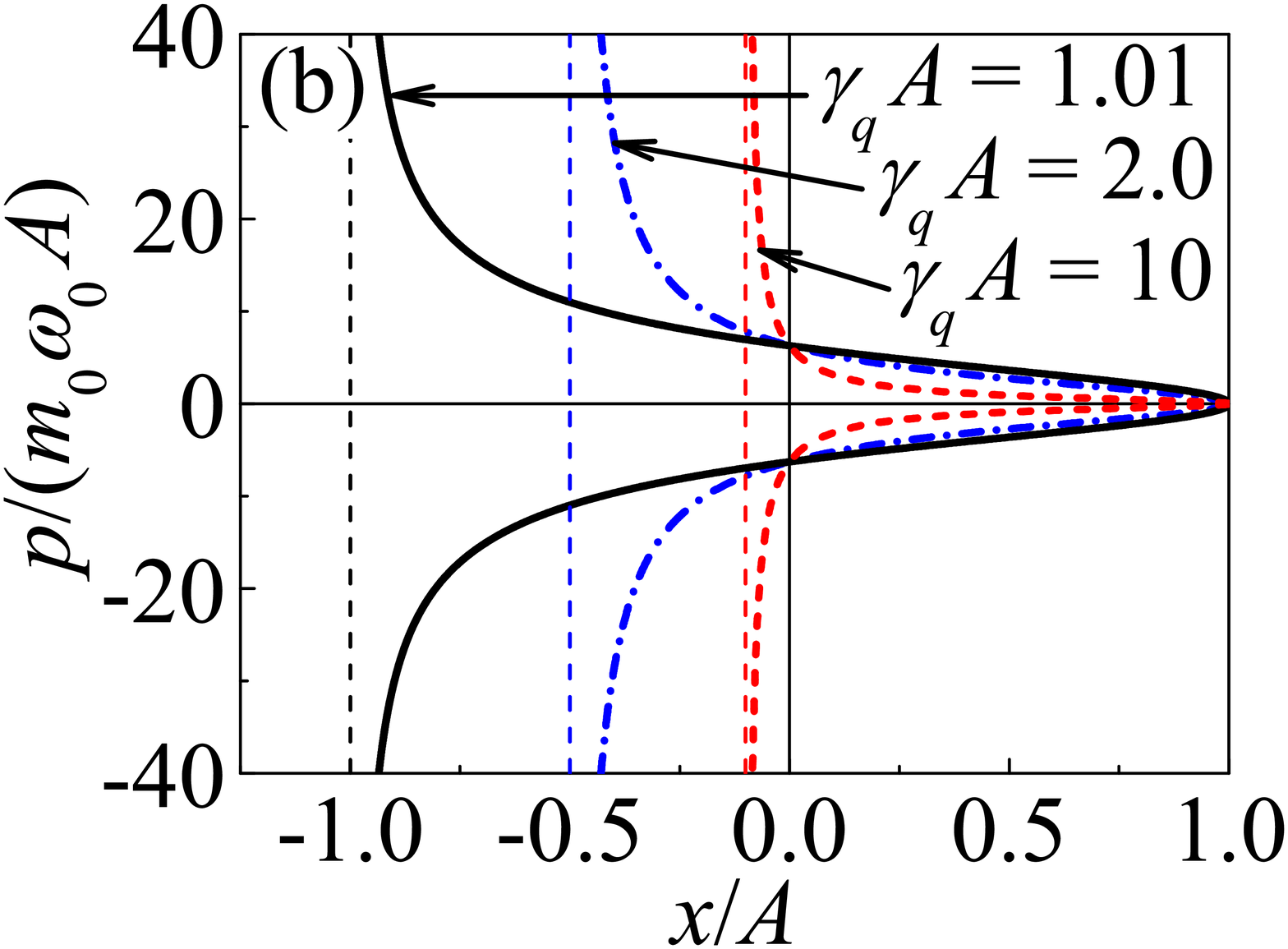}
 \end{minipage} \\
 \begin{minipage}[h]{0.49\linewidth}
  \includegraphics[width=\linewidth]{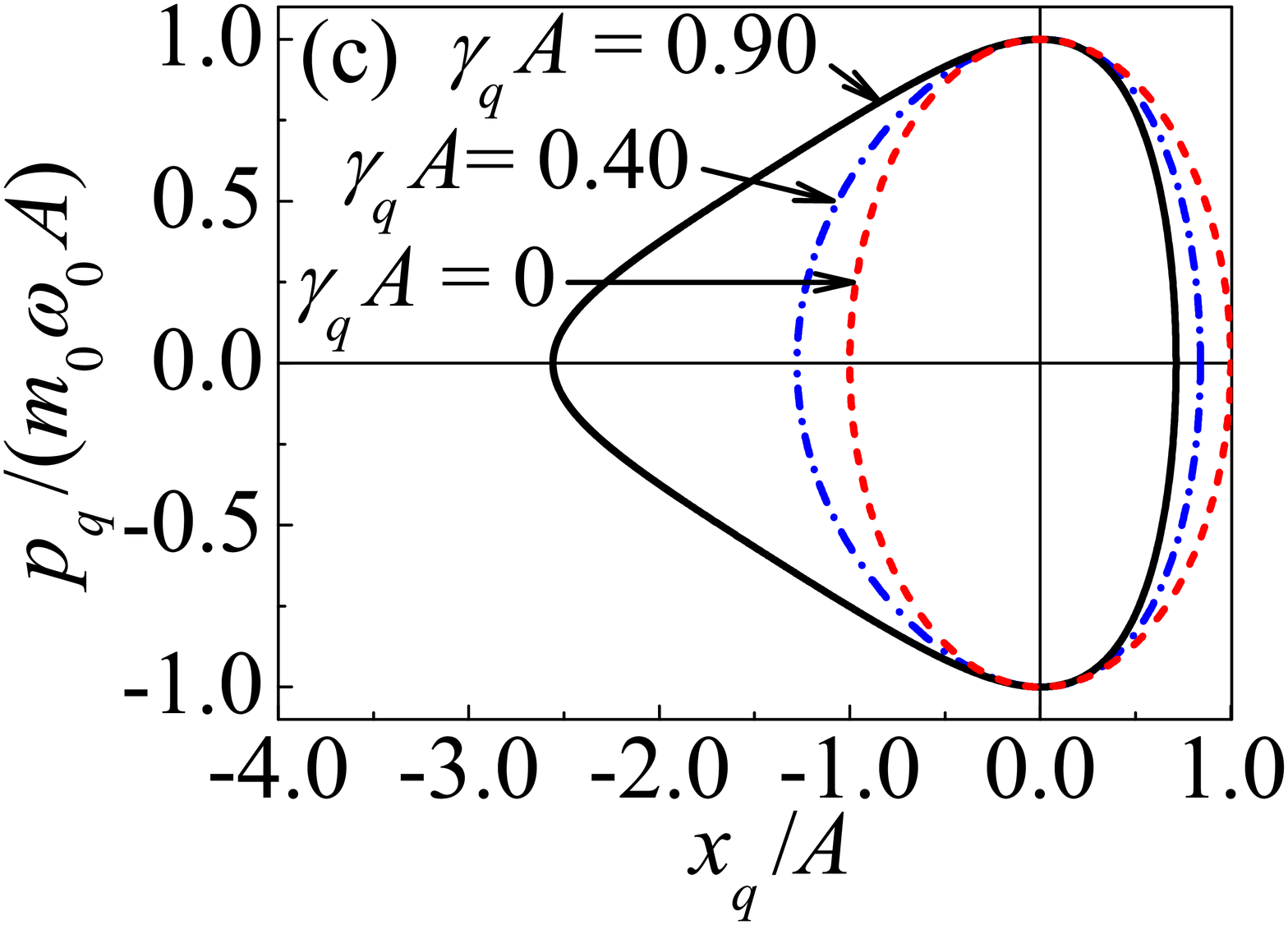}
 \end{minipage} 
 \begin{minipage}[h]{0.49\linewidth}
  \includegraphics[width=\linewidth]{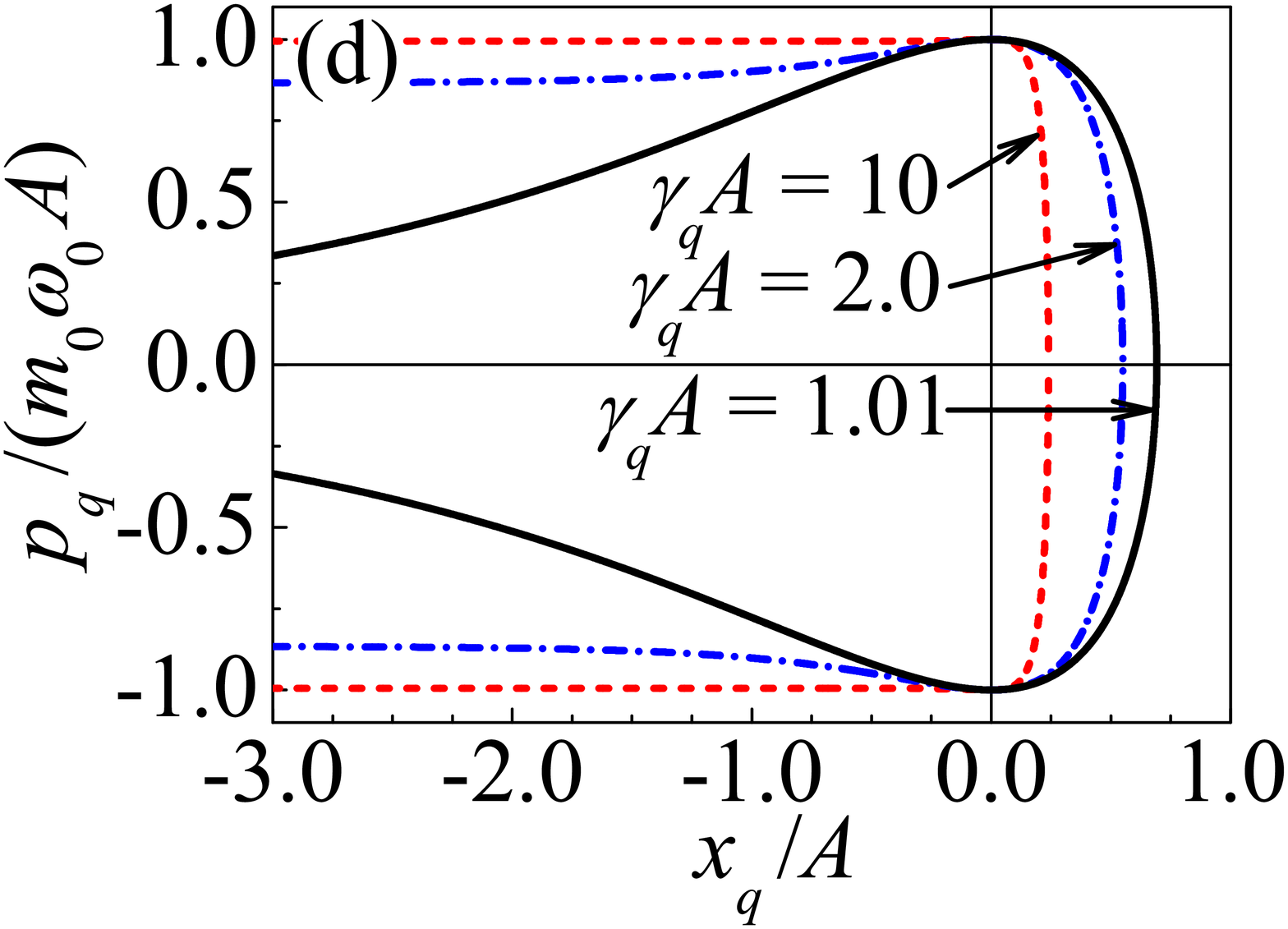}
 \end{minipage} 
 \caption{\label{fig:phase-space}
  (Color online) Phase spaces for a classical oscillator with 
  $m(x)$ given by Eq.~(\protect\ref{eq:m(x)}).
  Upper line: undeformed canonical coordinates $(x, p)$.
  Bottom line: deformed canonical coordinates $(x_q, p_q)$.
  Left column: closed orbits. Right column: open orbits.
 }
\end{figure}

\begin{figure}[!htb]
 \centering
 \begin{minipage}[h]{0.48\linewidth}
  \includegraphics[width=\linewidth]{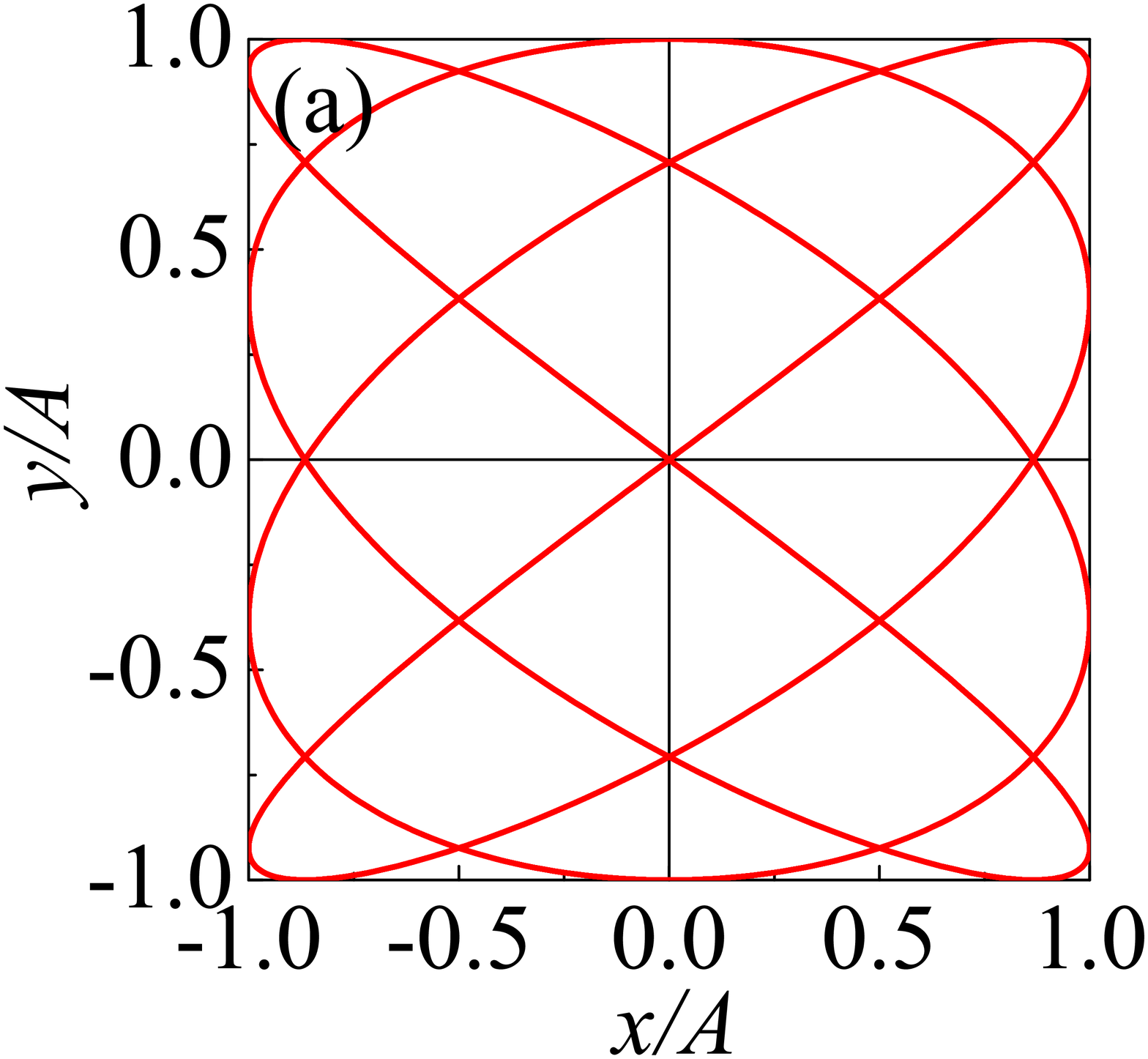}
 \end{minipage} 
 \begin{minipage}[h]{0.48\linewidth}
  \includegraphics[width=\linewidth]{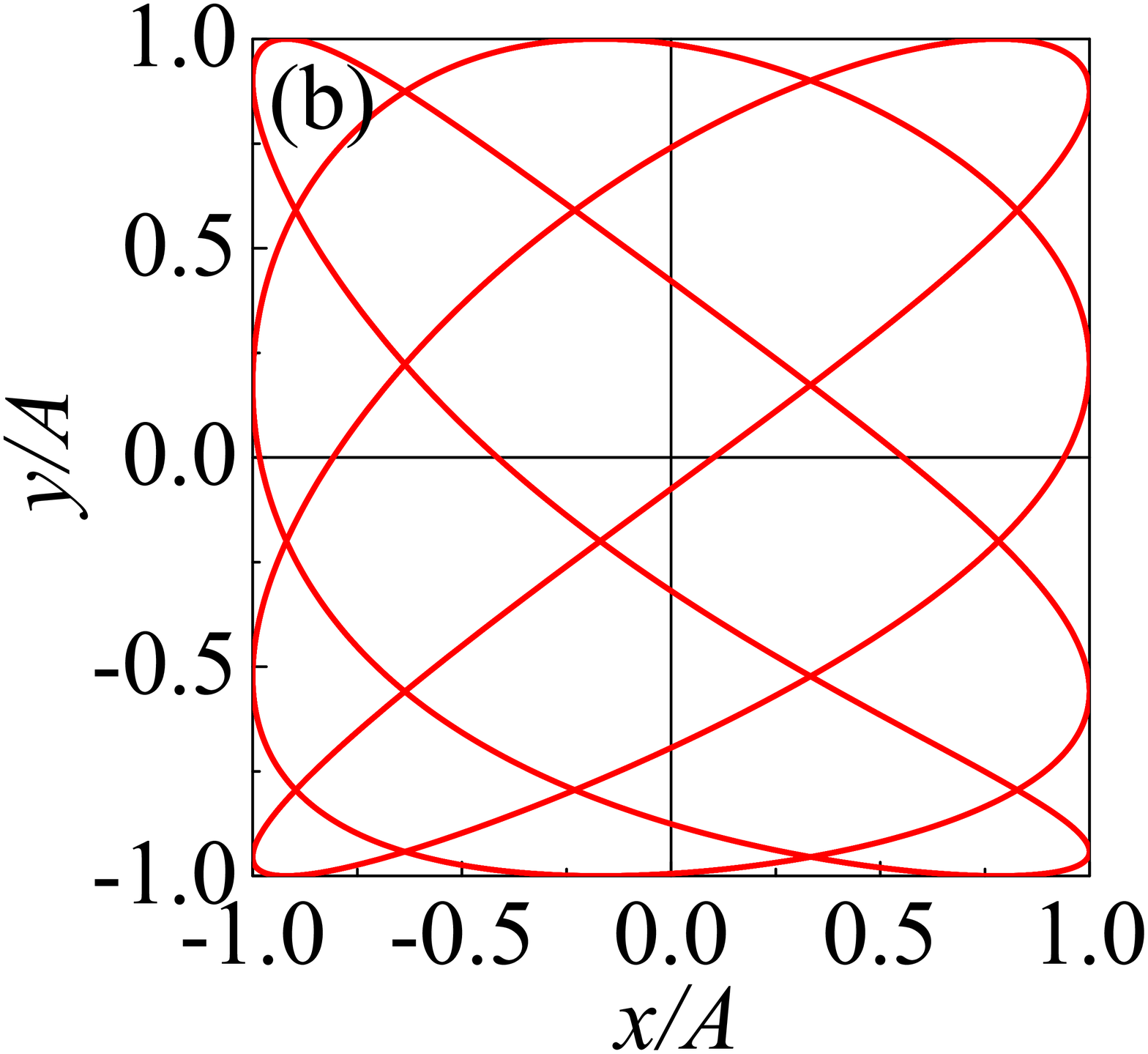}
 \end{minipage} \\
 \begin{minipage}[h]{0.48\linewidth}
  \includegraphics[width=\linewidth]{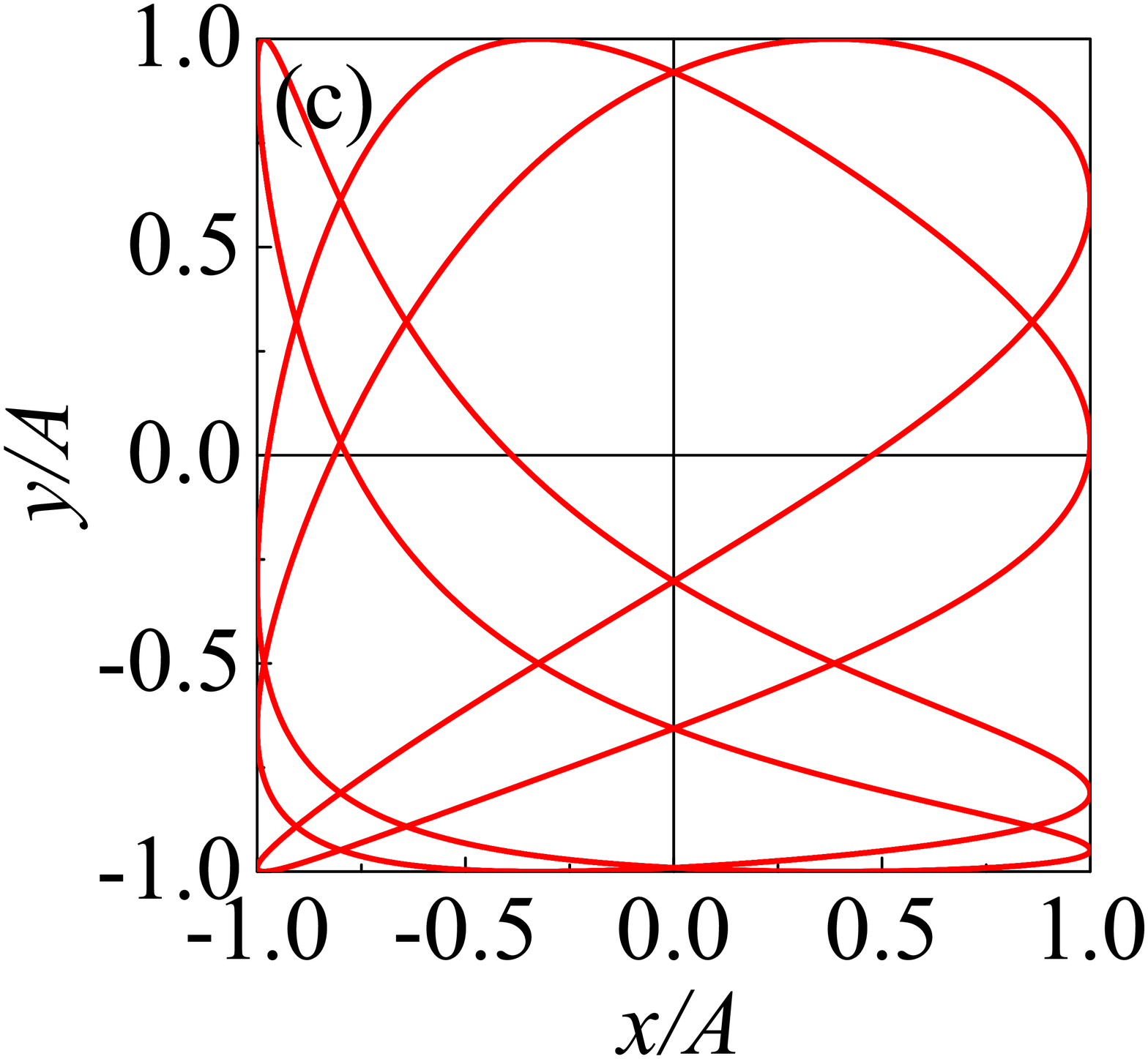}
 \end{minipage} 
 \begin{minipage}[h]{0.48\linewidth}
  \includegraphics[width=\linewidth]{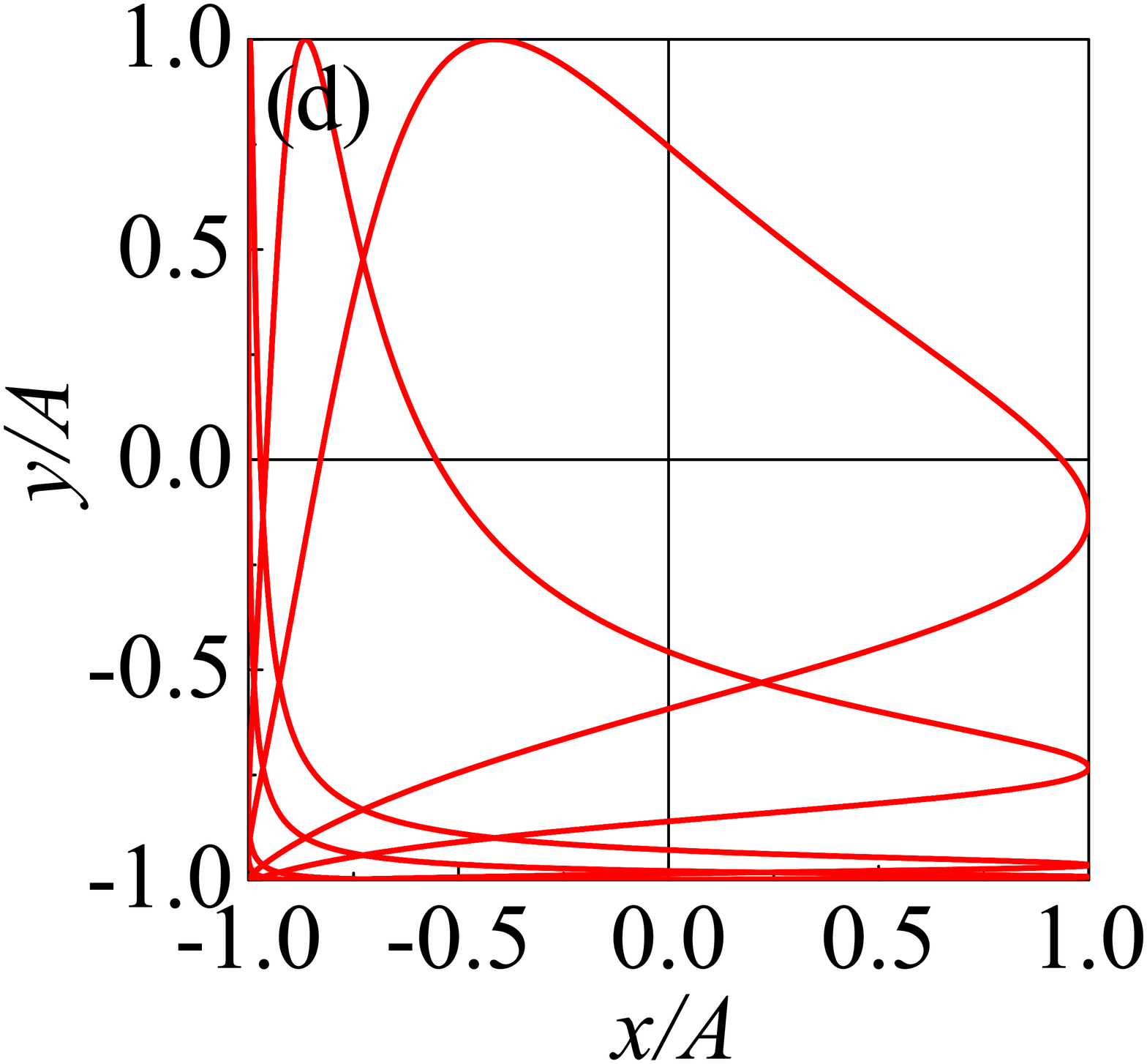}
 \end{minipage}
 \caption{\label{fig:lissajous} 
 (Color online)
 Lissajous figures for two deformed 
 oscillators with phase difference $\delta_y - \delta_x   = \pi/2$ and 
 $\omega_y/\omega_x = 4/3$ with 
 $\gamma_qA$ equal to (a) 0, (b) 0.3, (c) 0.5, and (d) 0.9.}
\end{figure}

Equation~(\ref{eq:canonical-transf}) transforms the Hamiltonian $H(x, p)$
to the Morse oscillator Hamiltonian \cite{morse}
\begin{equation}
\label{eq:morse} 
K(x_q, p_q) = \frac{p_q^2}{2m_0} + W_q(e^{-\alpha_q x_q}-1)^2,
\end{equation}
with the binding energy $W_q \equiv m_0\omega_0^2/2\gamma_q^2,$
and $\alpha_q \equiv - \gamma_q$ is a parameter that controls 
the anharmonicity of the potential
($\gamma_q^2 A^2 = E/W_q$).
Therefore, the classical canonical transformation 
Eq.~(\ref{eq:canonical-transf})
maps a system with PDM given by
Eq.~(\ref{eq:m(x)}) subject to a quadratic potential in a phase space $(x, p)$ 
into a system of constant mass subject to a Morse potential  in the 
deformed phase space $(x_q, p_q)$.
Since $\gamma_q^2 A^2 = E/W_q$, 
the particle presents closed curves in the phase space $(x_q, p_q)$ 
for $E < W_q$ and an open curves for $E > W_q$.
Figure \ref{fig:phase-space} shows the phase spaces
$(x, p)$ and $(x_q, p_q)$ for $\gamma_q A < 1$ (closed orbits)
and $\gamma_q A > 1$ (open orbits). 
The confined position
($x_{\text{min}} < x \le A$) 
and divergent momentum ($-\infty < p < \infty$)
for the problem with $\gamma_q A > 1$ 
is turned into unconfined deformed position
($-\infty < x_q < x_{q,\text{max}} = \gamma_q^{-1} \ln(1+\gamma_q A)$)
and bounded deformed momentum
($-p_{q,\text{max}} < p_q < p_{q,\text{max}}$, 
 with $p_{q,\text{max}} = m_0 \omega_0 A \sqrt{1-\frac{1}{\gamma_q^2 A^2 }}$).
Figure \ref{fig:lissajous} brings some instances of Lissajous 
curves for two oscillators.

The probability 
$ P_{\mbox{\footnotesize classic}}(x)dx \propto {dx}/{v} $
to find the particle with position between $x$ and $x + dx$ is
\begin{equation}
	\label{eq:probability_classic}
	P_{\text{classic}}(x) dx 
	= \frac{\sqrt{1 - \gamma_q^2 A^2}}
		 {\pi (1+\gamma_q x)\sqrt{A^2 - x^2}} dx.
\end{equation}
The first and second moments of position and momentum are
\begin{subequations}
\label{eq:x-p-x^2-p^2-medium}
\begin{equation}
\label{eq:x-medium}
\frac{\overline{x}}{A} 
   = -\frac{1 - \sqrt{1 - \gamma_q^{2} A^2}}{\gamma_q A},
\end{equation}

\begin{equation}
\label{eq:x^2-medium}
  \frac{\overline{x^2}}{A^2} 
   = \frac{1 - \sqrt{1 - \gamma_q^{2} A^2}}{\gamma_q^2 A^2},
\end{equation}
\begin{equation}
\label{eq:p-medium}
  \overline{p} = 0,
\end{equation}
\begin{equation}
\label{eq:p^2-medium}
  \overline{p^2} = 
 \frac{m_0^2 \omega_0^2 A^2}{2(1 - \gamma_q^2 A^2)}.
\end{equation}
\end{subequations}
The virial theorem is only valid for the usual case $\gamma_q = 0$
since (see Eq.~(\ref{eq:x^2-medium}), with $\overline{T} = E - \overline{V}$):
$\overline{T} = \sqrt{1 - \gamma_q^2 A^2}\; \overline{V}$.

The use of WKB approximation is an alternative and simple way 
to obtain the energy levels of a quantum system through its classical analogue.
This procedure leads to
\begin{eqnarray}
\left( n + \frac{1}{2} \right) \pi \hbar 
	& = &
	\displaystyle \frac{1}{2\pi}\int_{-A}^{A} pdx 
	= \frac{m \omega_0}{2\pi}\int_{-A}^{A}
	\frac{\sqrt{A^2 - x^2}}{1 + \gamma_q x}dx 
	\nonumber \\
	& = & 
	\displaystyle \frac{m\omega_0 A^2}{4\pi} 
	\int_{0}^{2\pi} \frac{\sin^2\theta_q}{1+\gamma_q A \cos \theta_q}d\theta_q
\end{eqnarray}
with $n$ integer.
Solving the above equation for $A^2 = 2E_n/m_0\omega^2_0$,
we arrive at
\begin{eqnarray}
\label{eq:levels-energy}
E_n & = & \omega_0 {\hbar} \left( n + \frac{1}{2} \right)
          \left[
           1-\frac{\gamma_q^2 x_0^2}{2} \left( n + \frac{1}{2} \right) 
          \right],
\end{eqnarray}
where $x_0^2 = \hbar/m_0 \omega_0$.
Ref.\ \cite{costa-filho-2013} obtained the same result
by using a non Hermitian linear momentum,
Eq.~(\ref{eq:operator-momentum-generalized}).
The energy levels of the quantum harmonic oscillator 
with PDM
given by Eq.~(\ref{eq:m(x)}) are identical to those 
of a constant mass particle in a constant Morse potential, 
since these two systems may be mapped into one another 
by  the canonical transform (\ref{eq:canonical-transf}).

\begin{figure}[!htb]
 \centering
 \begin{minipage}[h]{0.48\linewidth}
  \includegraphics[width=\linewidth]{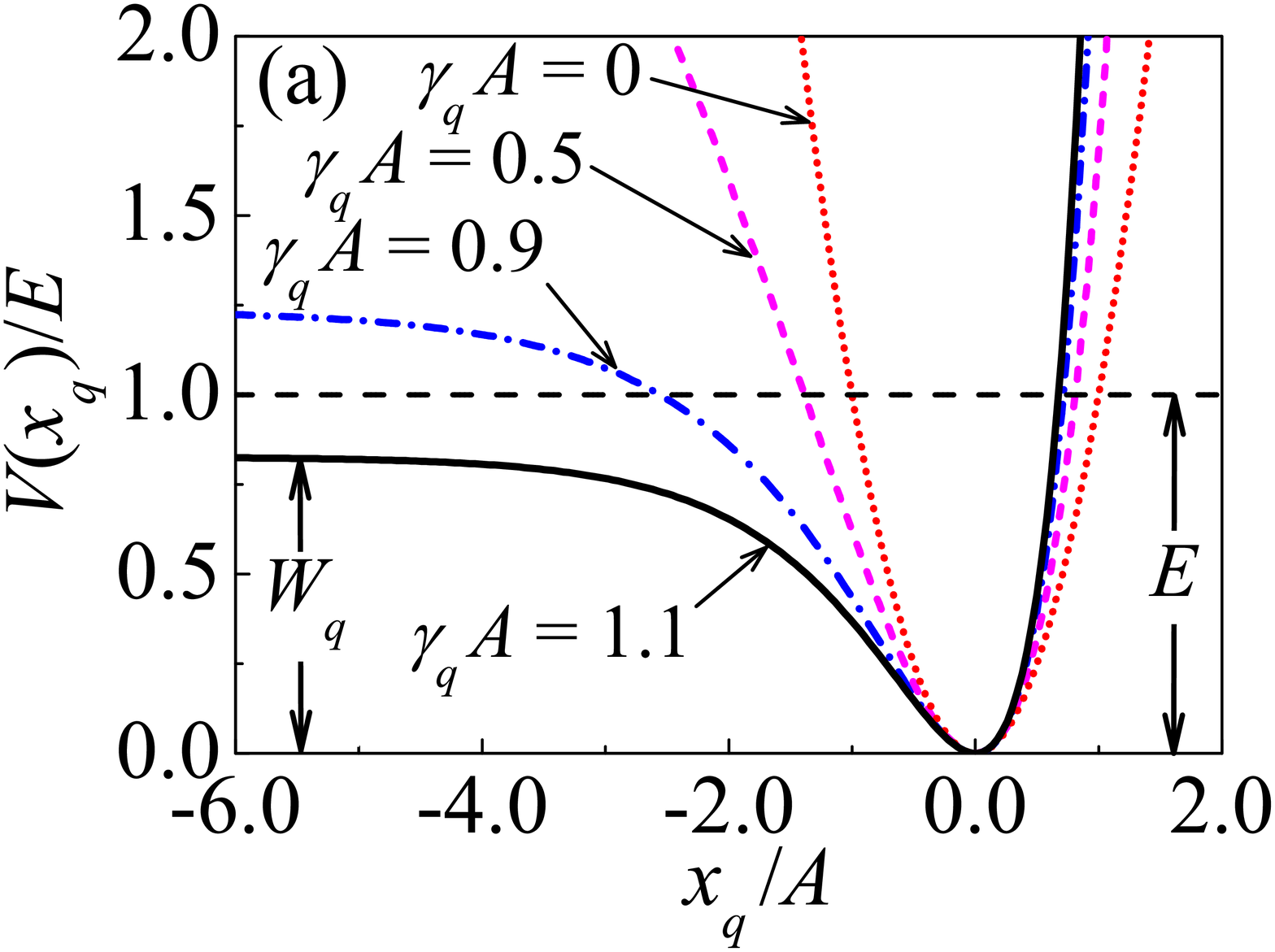}
 \end{minipage}
 \begin{minipage}[h]{0.48\linewidth}
  \includegraphics[width=\linewidth]{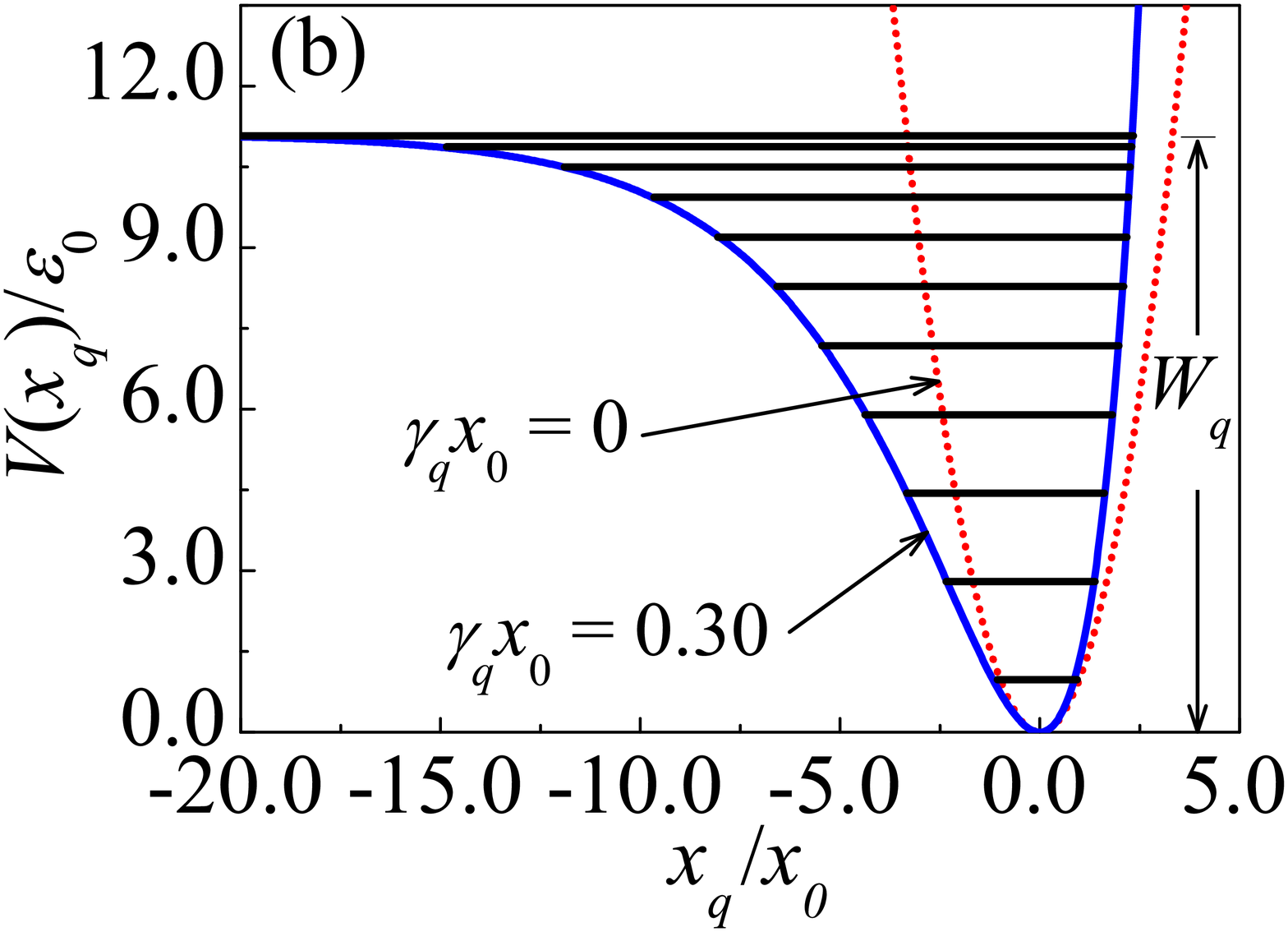}
 \end{minipage}
 \caption{\label{fig:potentials} 
  (Color online)
  (a) Potentials in the phase space and $(x_q, p_q)$ 
  for values of $\gamma_qA =$ 
  0 (dotted red), 0.5 (dashed magenta), 
  0.9 (dash-dotted blue), and  1.1 (solid black).
  (b) Energy levels of the bound states for $\gamma_q x_0 = 0.3$,
  with $\varepsilon_0 = \hbar \omega_{0}/2$.
  The usual case $\gamma_q x_0=0$ is shown for comparison.
 }
\end{figure}

Figure \ref{fig:potentials} shows the
Morse potential $V(x_q) = W_q(e^{-\alpha_q x_q}-1)^2$ 
for different values of $\gamma_q A$, as well as 
energy levels of bound states.

\section{Quantum harmonic oscillator with position-dependent mass}

\begin{figure}[!htb]
 \begin{minipage}[h]{0.75\textwidth}
 \centering
 \begin{minipage}[h]{0.48\textwidth}
  \includegraphics[width=\textwidth]{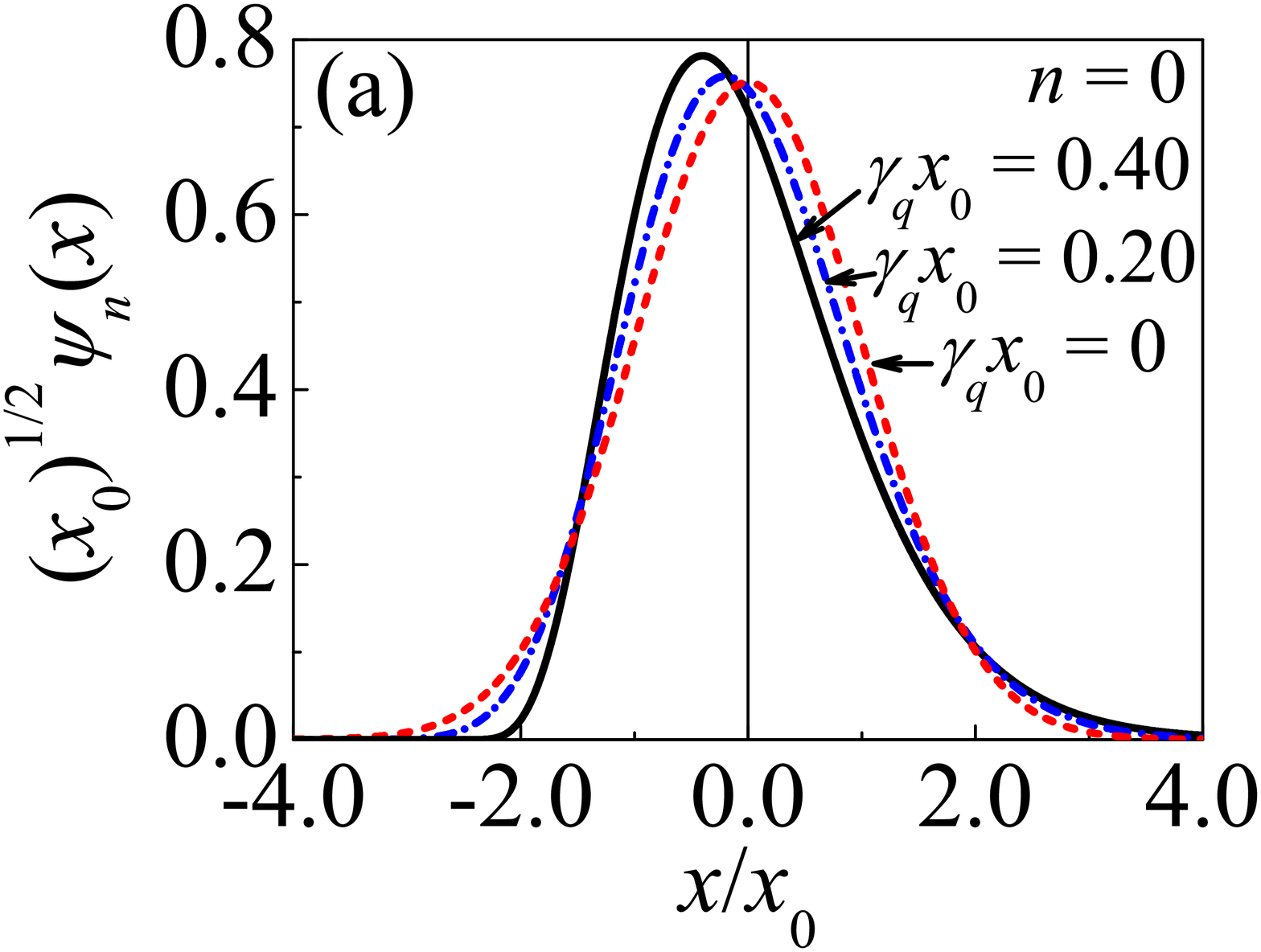}
 \end{minipage} 
 \begin{minipage}[h]{0.48\textwidth}
  \includegraphics[width=\textwidth]{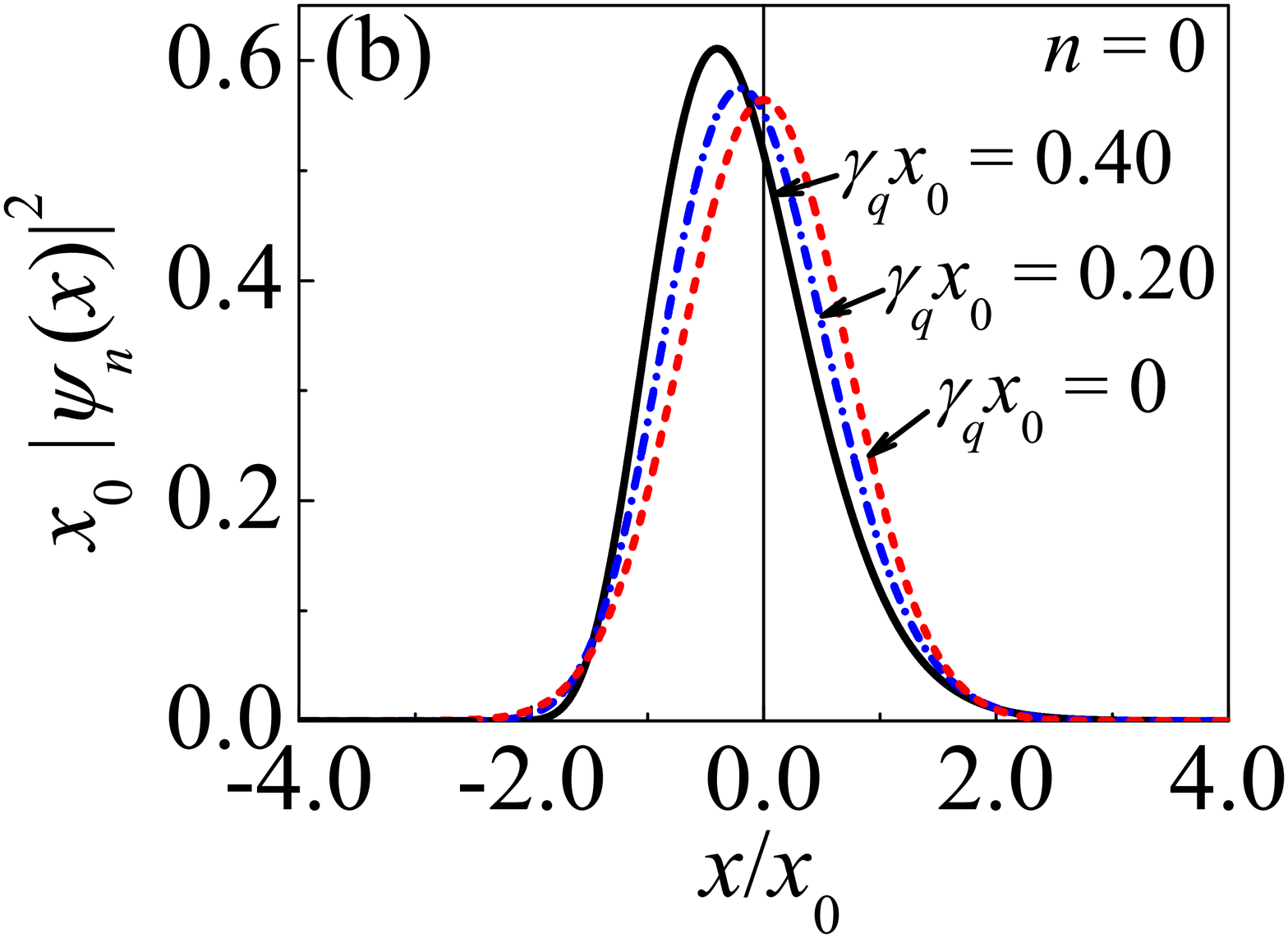}
 \end{minipage}\\ 
 \begin{minipage}[h]{0.48\textwidth}
  \includegraphics[width=\textwidth]{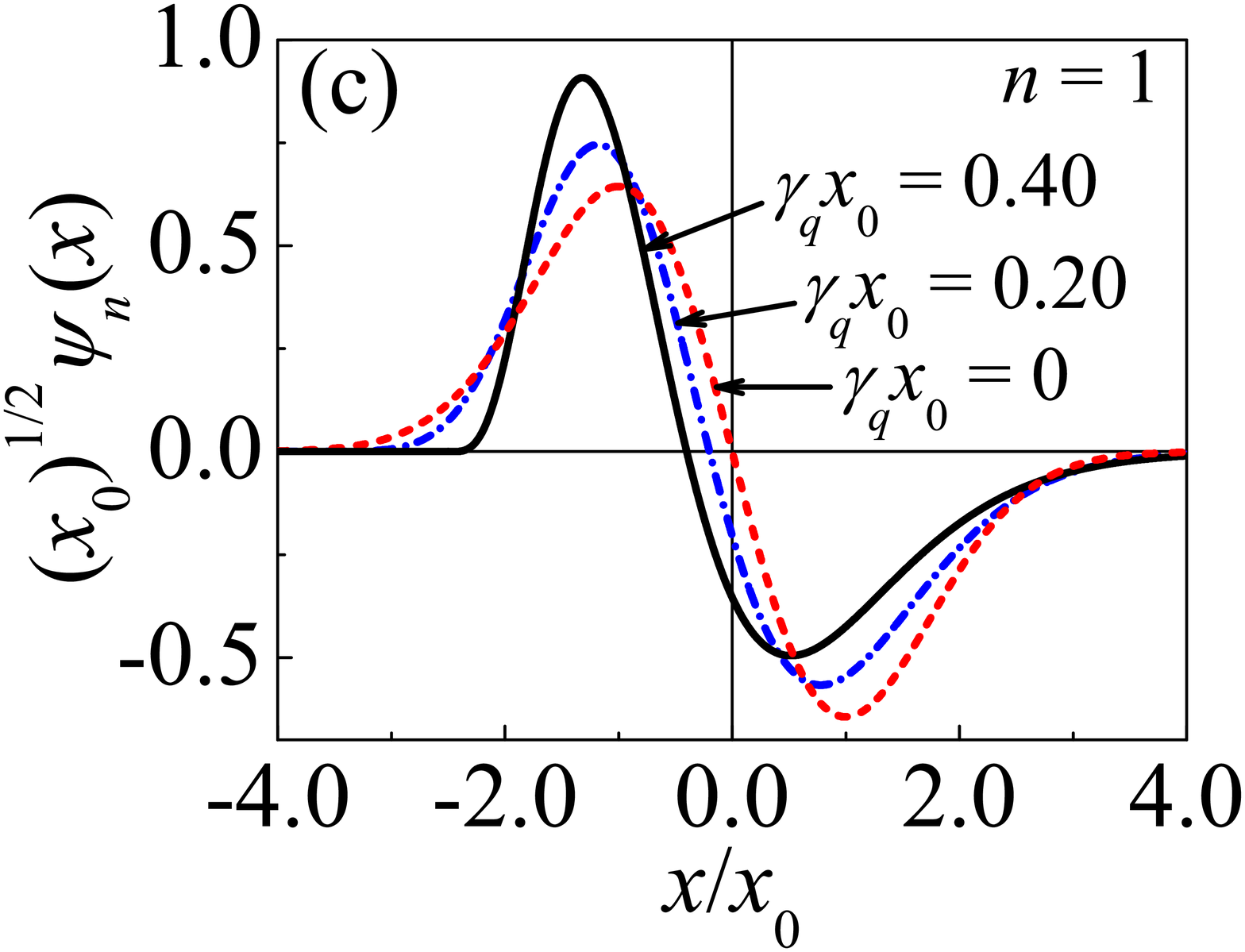}
 \end{minipage} 
 \begin{minipage}[h]{0.48\textwidth}
  \includegraphics[width=\textwidth]{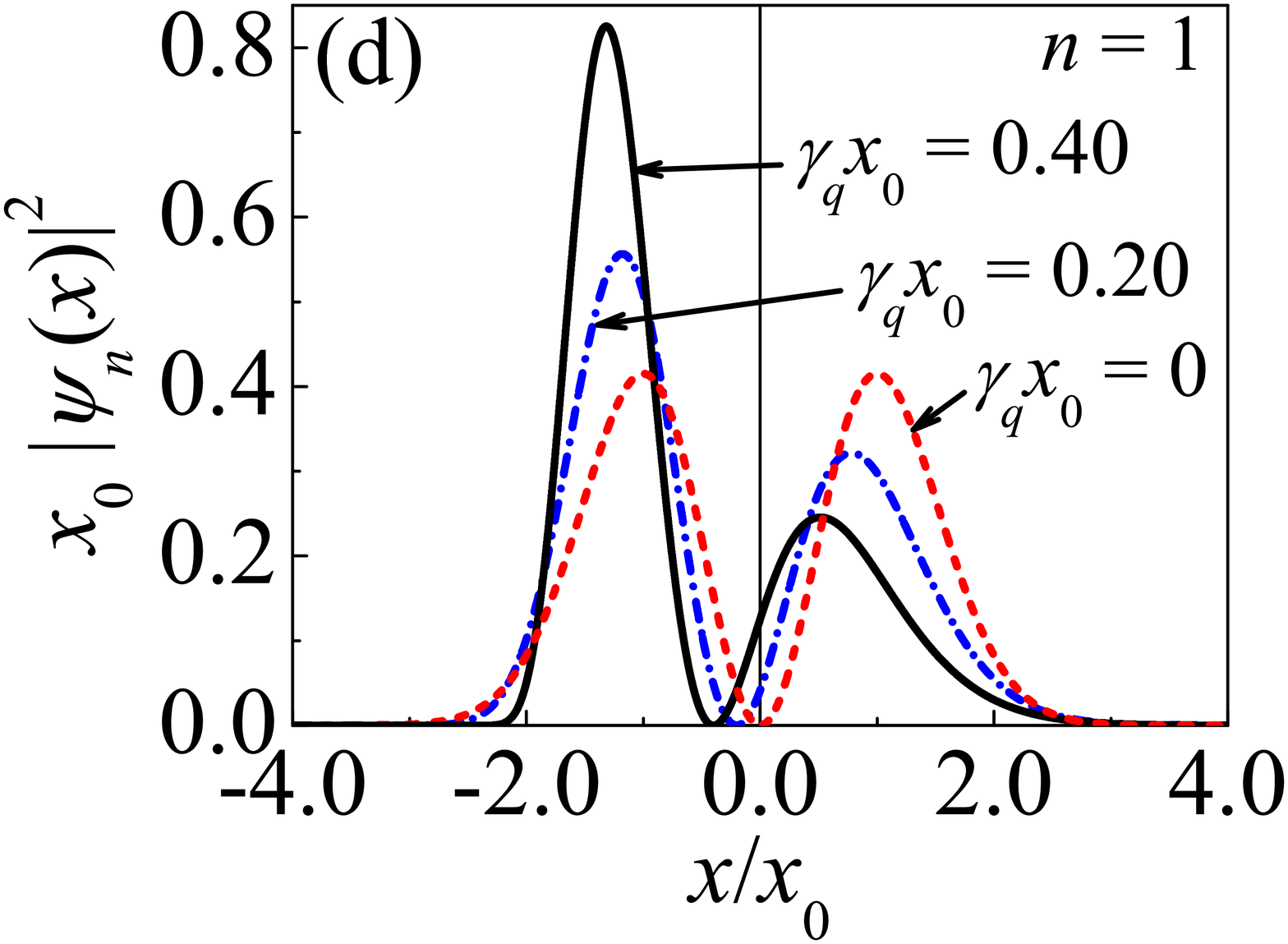}
 \end{minipage} \\
 \begin{minipage}[h]{0.48\textwidth}
  \includegraphics[width=\textwidth]{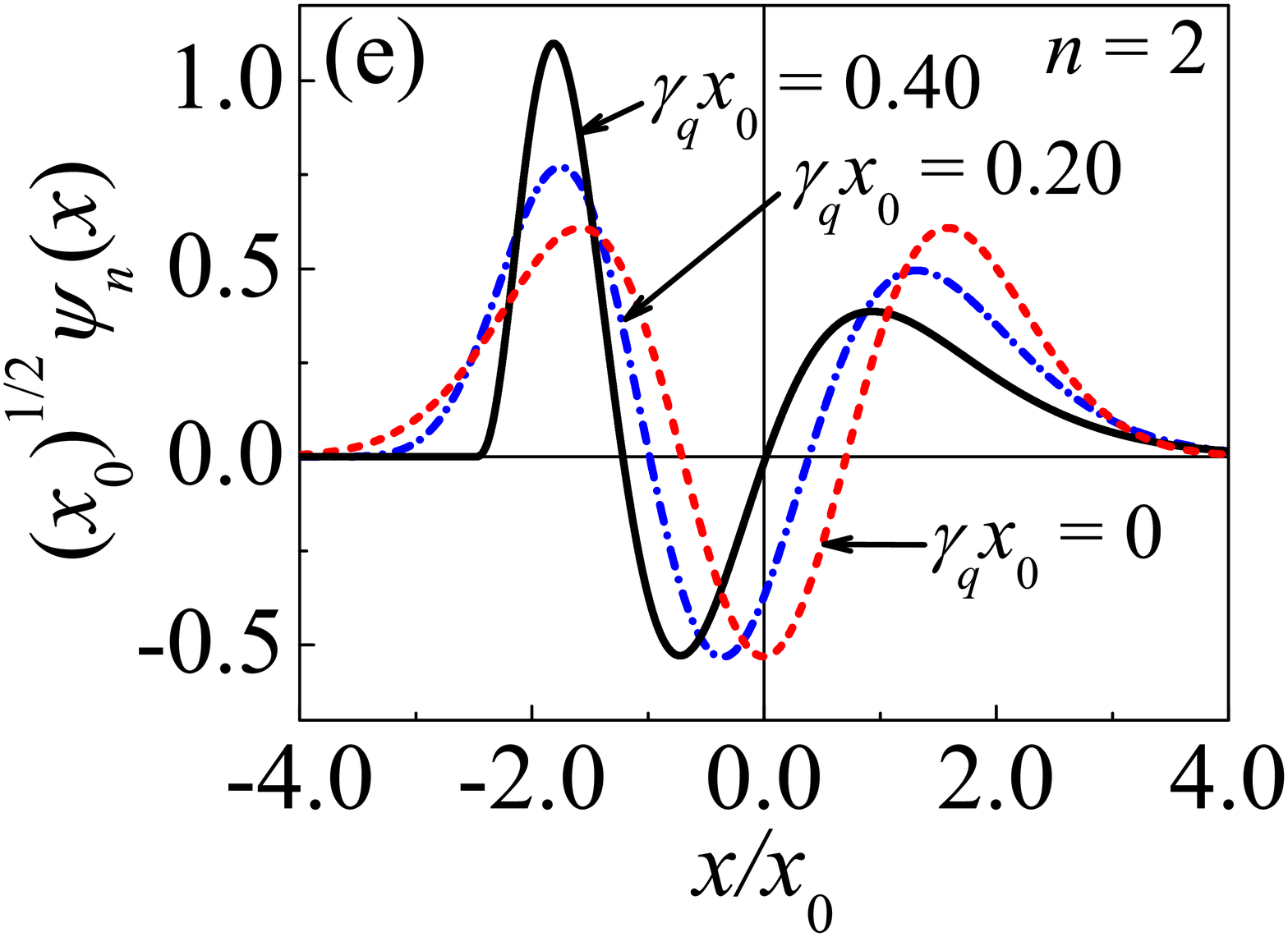}
 \end{minipage}
 \begin{minipage}[h]{0.48\textwidth}
  \includegraphics[width=\textwidth]{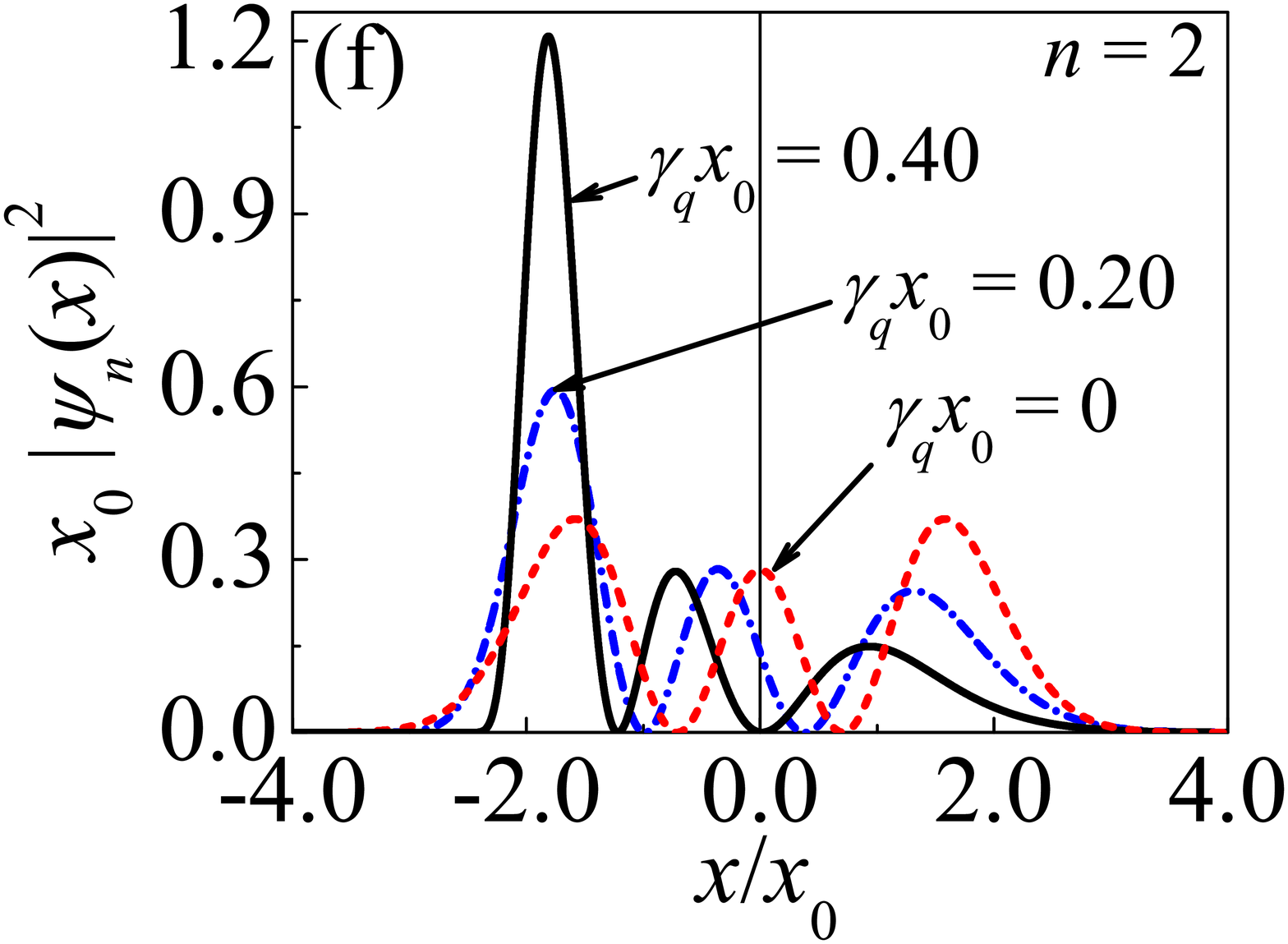}
 \end{minipage}\\
 \begin{minipage}[h]{0.48\textwidth}
  \includegraphics[width=\textwidth]{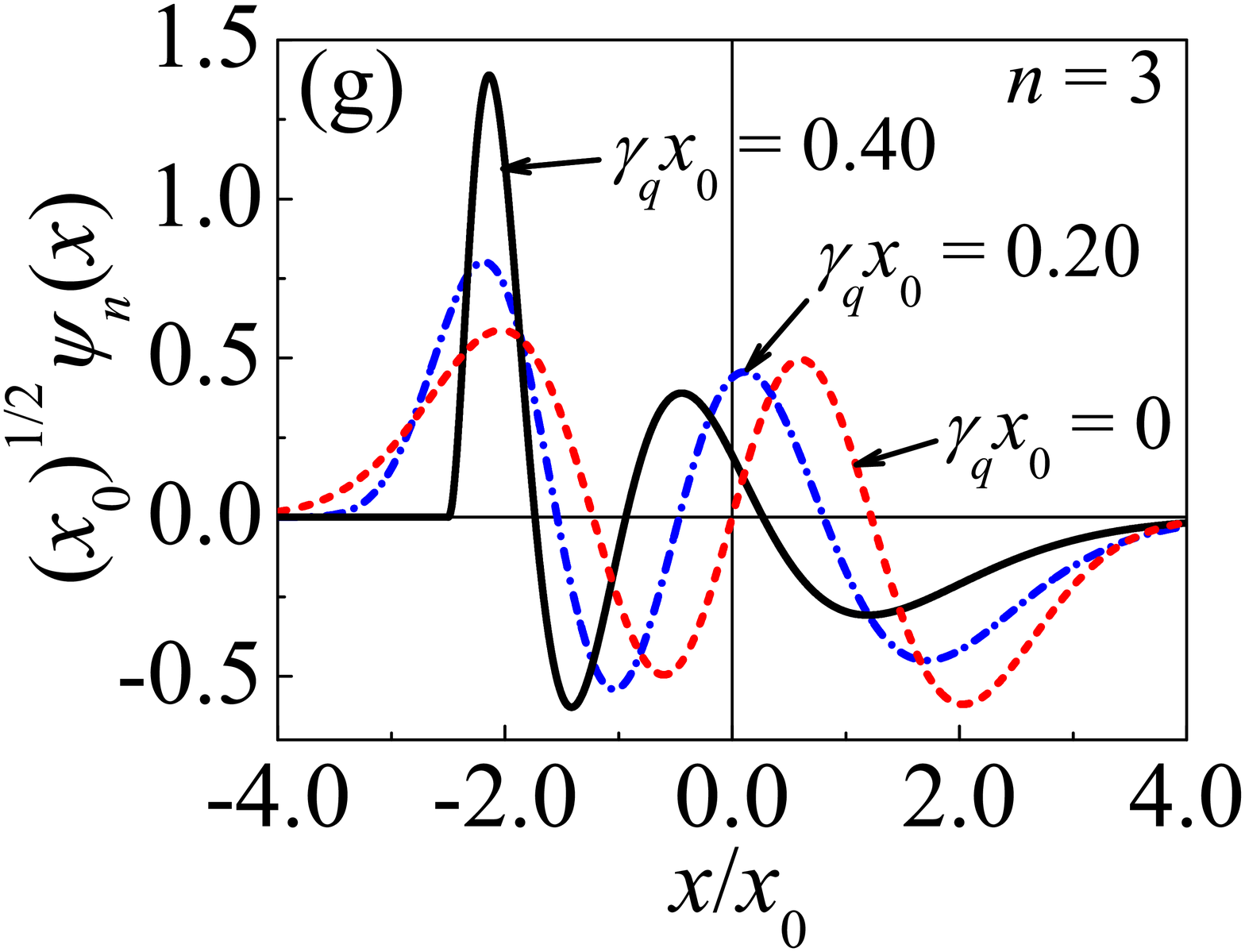}
 \end{minipage}
 \begin{minipage}[h]{0.48\textwidth}
  \includegraphics[width=\textwidth]{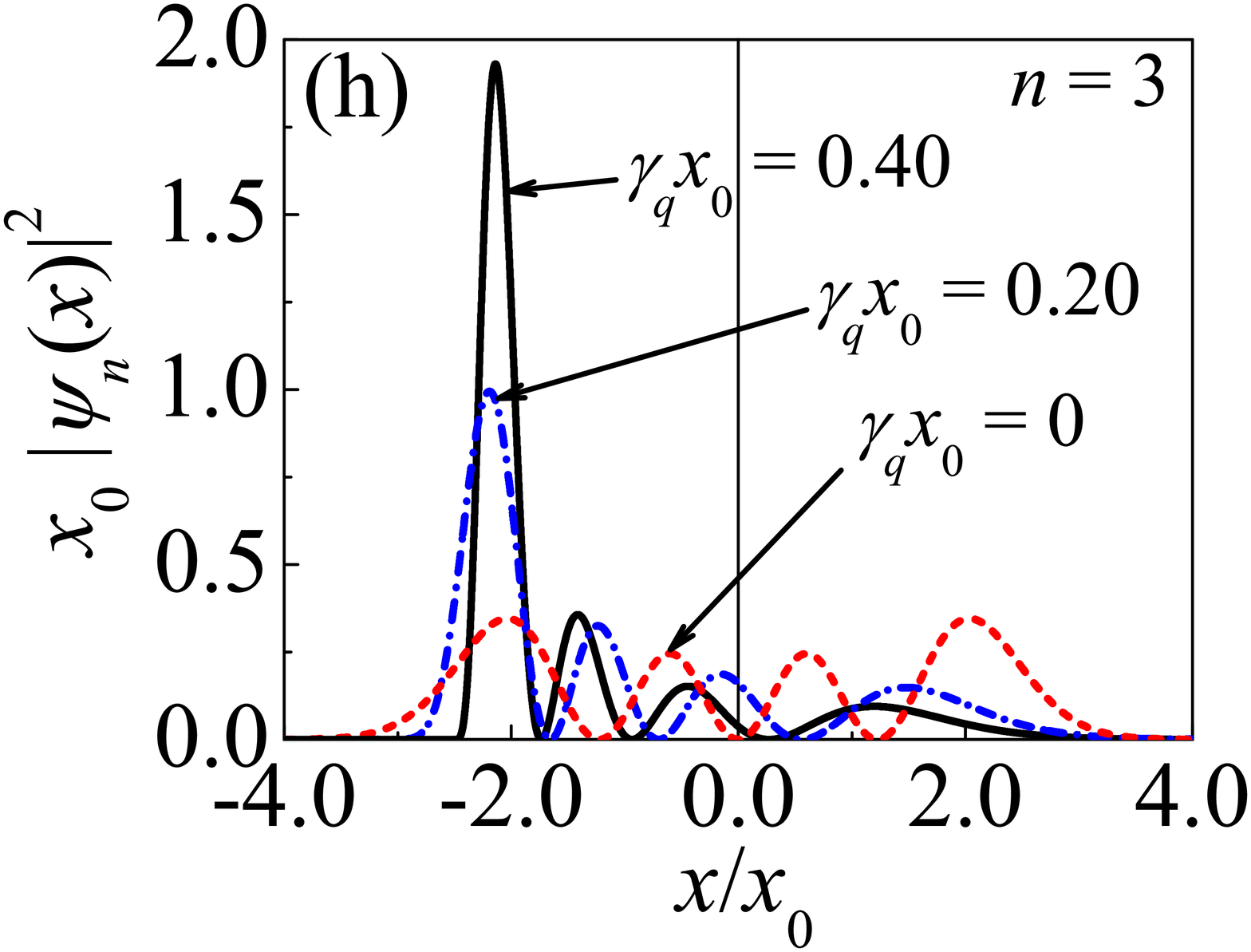}
 \end{minipage}
 \end{minipage}
 \caption{\label{fig:psi_and_psi_quad} 
  (Color online)
  Wave functions $\psi_n (x)$ (left column)
  and probability densities $|\psi_n (x)|^2 $ (right column)
  for a particle with PDM
  according to Eq.~(\protect\ref{eq:m(x)})
  under a quadratic potential for different values of $\gamma_q x_0$
  (indicated, usual case $\gamma_q x_0 = 0$ is shown, for comparison).
  (a) and (b): $n=0$ (ground state), 
  (c) and (d): $n=1$ (first excited state),
  (e) and (f): $n=2$ (second excited state).
  (g) and (h): $n=3$ (third excited state).
}
\end{figure}

\begin{figure}[!htb]
\centering
\includegraphics[width=0.65\linewidth ]{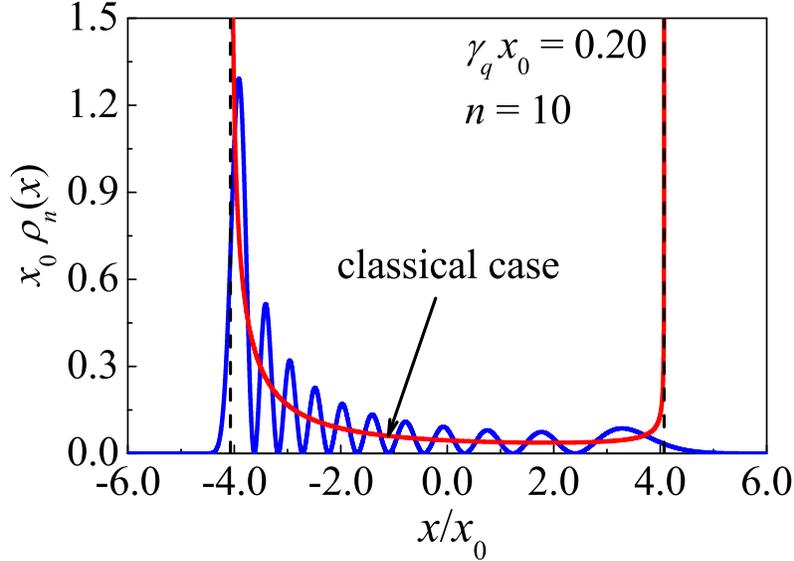}
\caption{\label{fig:correspondencia} 
(Color online)
Probability density for a PDM
(Eq.~(\protect\ref{eq:m(x)})) in a harmonic potential for 
$x_0 \gamma_q = 0.20$ and $n=10$.
The classical case (Eq.~(\protect\ref{eq:probability_classic}))
is  represented for comparison.
}
\end{figure}

The Hamiltonian operator for the PDM oscillator is
\begin{equation}   
\label{eq:hamiltonian-pdm-general}
    \displaystyle \hat{H} 
	= \displaystyle -\frac{\hbar^2}{2} 
	\biggl\{
	\displaystyle{
	[m(\hat{x})]^{-1/4} 
	\frac{d}{dx} [m(\hat{x})]^{-1/2} 
	\frac{d}{dx} [m(\hat{x})]^{-1/4}
	}
	\biggr\}
	\displaystyle + \frac{1}{2}m_0 \omega_0^2 \hat{x}^2,
\end{equation}
and its associated time independent Schr\"odinger equation at basis 
$\{ | x \rangle \}$ is
\begin{equation}
 \label{eq:schroedinger-equation-sho-pdm3}
  -\frac{{\hbar}^2 ( 1+\gamma_q x)^2}{2m_0} \frac{d^2\psi(x)}{dx^2}
  -\frac{{\hbar}^2\gamma_q( 1+\gamma_q x)}{m_0}\frac{d\psi(x)}{dx}
  -\frac{{\hbar}^2\gamma_{q}^2}{8m_0}\psi(x)
   + \frac{1}{2}m_0 \omega_0^2 x^2 \psi(x) = E\psi(x),
\end{equation}
where 
$\Psi(x, t) = \psi(x) e^{-iEt/\hbar}$ and
$\langle x | \alpha \rangle = \psi(x)$.
The the canonical transformation given by 
Eqs.~(\ref{eq:hermitian-operator-momentum-generalized}) 
and (\ref{eq:operator-position-generalized}) 
leads to the time independent Schr\"odinger equation at basis 
$\{ | x \rangle \}$
for a particle with constant mass $m_0$ submitted to the Morse potential:
\begin{equation}   
 -\displaystyle \frac{{\hbar}^2}{2m_0} \frac{d^2\varphi(x_q)}{dx_q^2}
 + W_{q}(e^{-\alpha_q x_q}-1)^2\varphi(x_q) 
 = E\varphi(x_q).
\end{equation} 
The corresponding time independent deformed Schr\"odinger equation for 
the state functions 
$\varphi_q(x) = \varphi (x_q(x)) = \sqrt{1 + \gamma_q x} \psi(x)$ is
\begin{equation}
\label{eq:deformed-schrodinger-equation-osc}
	-\frac{\hbar^2}{2m_0} D_{\gamma_q}^2 \varphi_q (x) 
	+ \frac{1}{2}m_0 \omega_0^2 x^2 \varphi_q (x)= E\varphi_q (x).
\end{equation} 
The solution of the above equations lead to the following wave functions:
\begin{equation}
\psi_n (x) = \frac{A_n}{\sqrt{1 + \gamma_q x}}
	e^{-d(1 + \gamma_q x)} [2d(1 + \gamma_q x)]^{b/2} 
	L_n^{(b)}(2d(1 + \gamma_q x)),
\end{equation}
where $d = 1/\gamma_q^2x_0^2$,
$b = 2d - 1 - 2n > 0$, $A_n^2 = b \gamma_q n!/(n+b)!$
and $L_n^{(b)}$ are associated Laguerre polynomials. 
The energy levels are also given by Eq.~(\ref{eq:levels-energy}).
Figure \ref{fig:psi_and_psi_quad} shows the wave function
and density probability. 
Figure \ref{fig:correspondencia} illustrates that for large quantum numbers, 
here exemplified with $n = 10$, the average of the quantum probability density 
$\rho_n(x) = |\psi_n (x)|^2$ approaches to the classical one given by 
Eq.~(\ref{eq:probability_classic}) with amplitude $A = a_{q,n}$, 
where
\begin{equation}
	a_{q,n}^2 = \frac{2E_n}{m_0 \omega_0^2} = x_0^2 (2n+1) 
	\left[ 1 - \frac{\gamma_q^2 x_0^2}{4} (2n+1) \right].
\end{equation}

The deformed translation operator (\ref{eq:translate}) 
can be extended to quantum systems 
with higher spatial dimensions through
$ 
\hat{T}_{q} (\vec{\varepsilon})|\vec{r}\,\rangle \equiv 
| x + (1+\gamma_{q} x)\varepsilon_x,  
y + (1+\gamma_{q} y)\varepsilon_y, 
z + (1+\gamma_{q} z)\varepsilon_z \rangle, 
$
with
$|\vec{r}\,\rangle = |x,y,z\rangle = |x\rangle \otimes |y\rangle \otimes |z\rangle $.
The linear momentum operator, 
Eq.\ (\ref{eq:hermitian-operator-momentum-generalized}),
is straightforwardly rewritten as
$\vec{p}_q = p_{q,x}\vec{\imath} + p_{q,y}\vec{\jmath} + p_{q,z}\vec{k}$, 
with 
\begin{equation}
\label{eq:generalized-momentum-3D}
\begin{array}{l}
\displaystyle{
 \hat{p}_{q,x} =  \frac{(\hat{1}+\gamma_q \hat{x})\hat{p}_x}{2} + 
                  \frac{\hat{p}_x(\hat{1}+\gamma_q \hat{x})}{2},
}	\\
\displaystyle{
 \hat{p}_{q,y} =  \frac{(\hat{1}+\gamma_q \hat{y})\hat{p}_y}{2} + 
                  \frac{\hat{p}_y(\hat{1}+\gamma_q \hat{y})}{2},
}	\\
\displaystyle{
 \hat{p}_{q,z} =  \frac{(\hat{1}+\gamma_q \hat{z})\hat{p}_z}{2} + 
                  \frac{\hat{p}_z(\hat{1}+\gamma_q \hat{z})}{2}.
}
\end{array}
\end{equation}
The canonical transformation (\ref{eq:canonical-transf}) 
extended for three dimensions becomes
$(\vec{p}, \vec{r}) \rightarrow (\vec{p}_q, \vec{r}_q)$
where
$\vec{r}_q = \gamma_q^{-1} 
             [  \ln(1+\gamma_q x)\vec{i} 
              + \ln(1+\gamma_q y)\vec{j} 
              + \ln(1+\gamma_q z)\vec{k}]$.

\begin{figure}[t!]
 \begin{minipage}[h]{0.90\textwidth}
  \centering
  \begin{minipage}[h]{0.45\linewidth}
   \includegraphics[width=\linewidth]{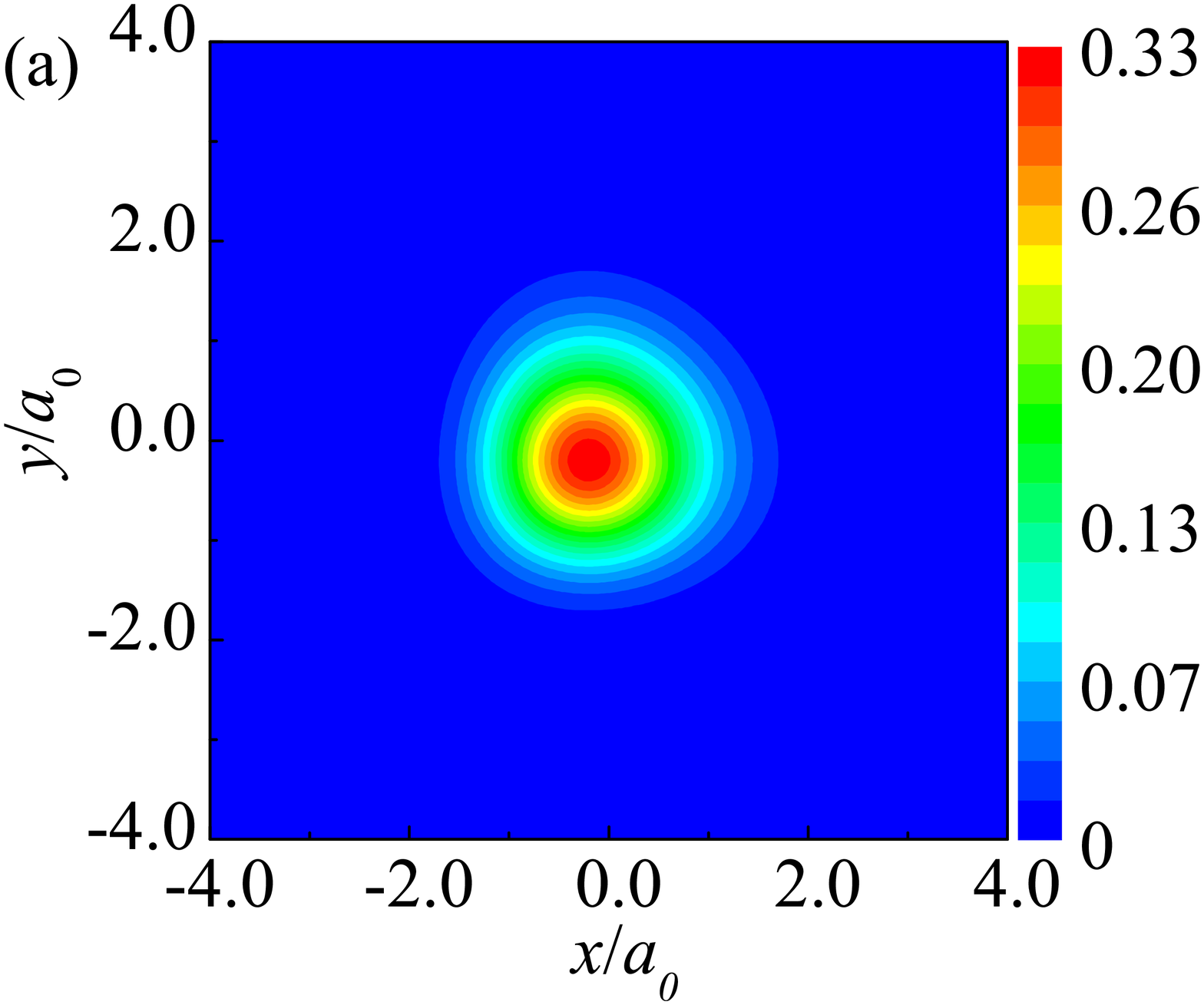}
  \end{minipage}
  \begin{minipage}[h]{0.45\linewidth}
   \includegraphics[width=\linewidth]{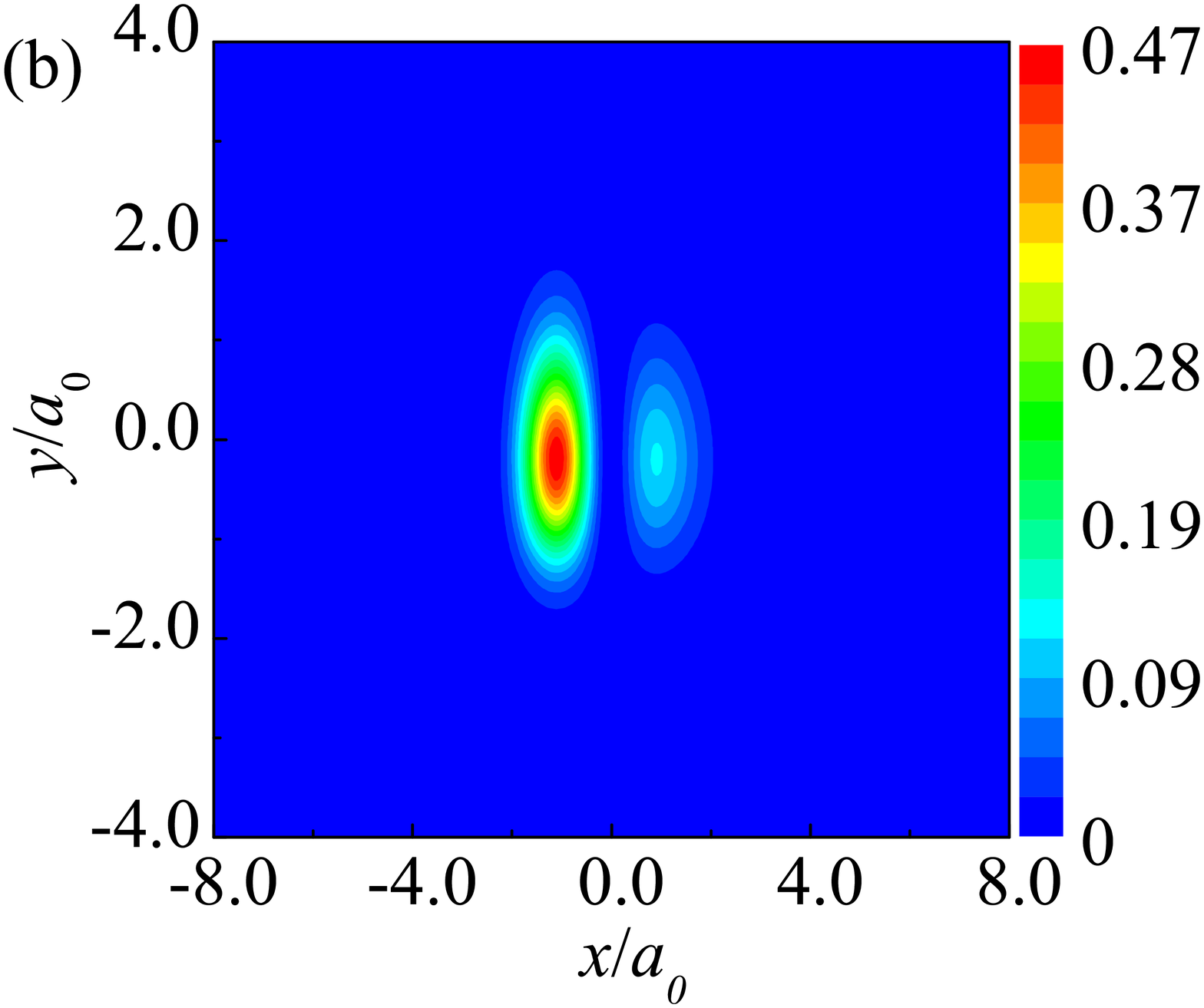}
  \end{minipage}
  \\
  \begin{minipage}[h]{0.45\linewidth}
   \includegraphics[width=\linewidth]{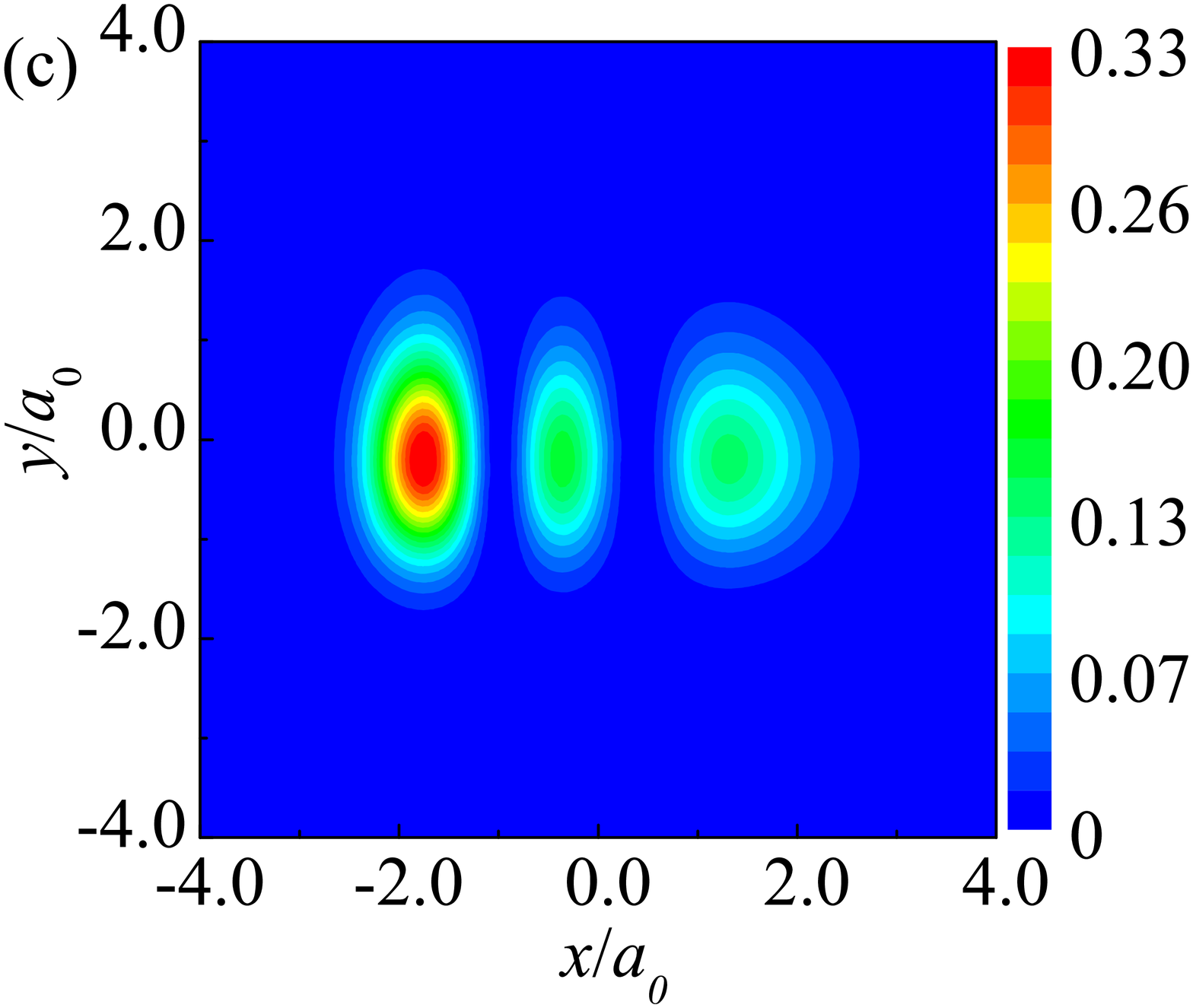}
  \end{minipage}
  \begin{minipage}[h]{0.45\linewidth}
   \includegraphics[width=\linewidth]{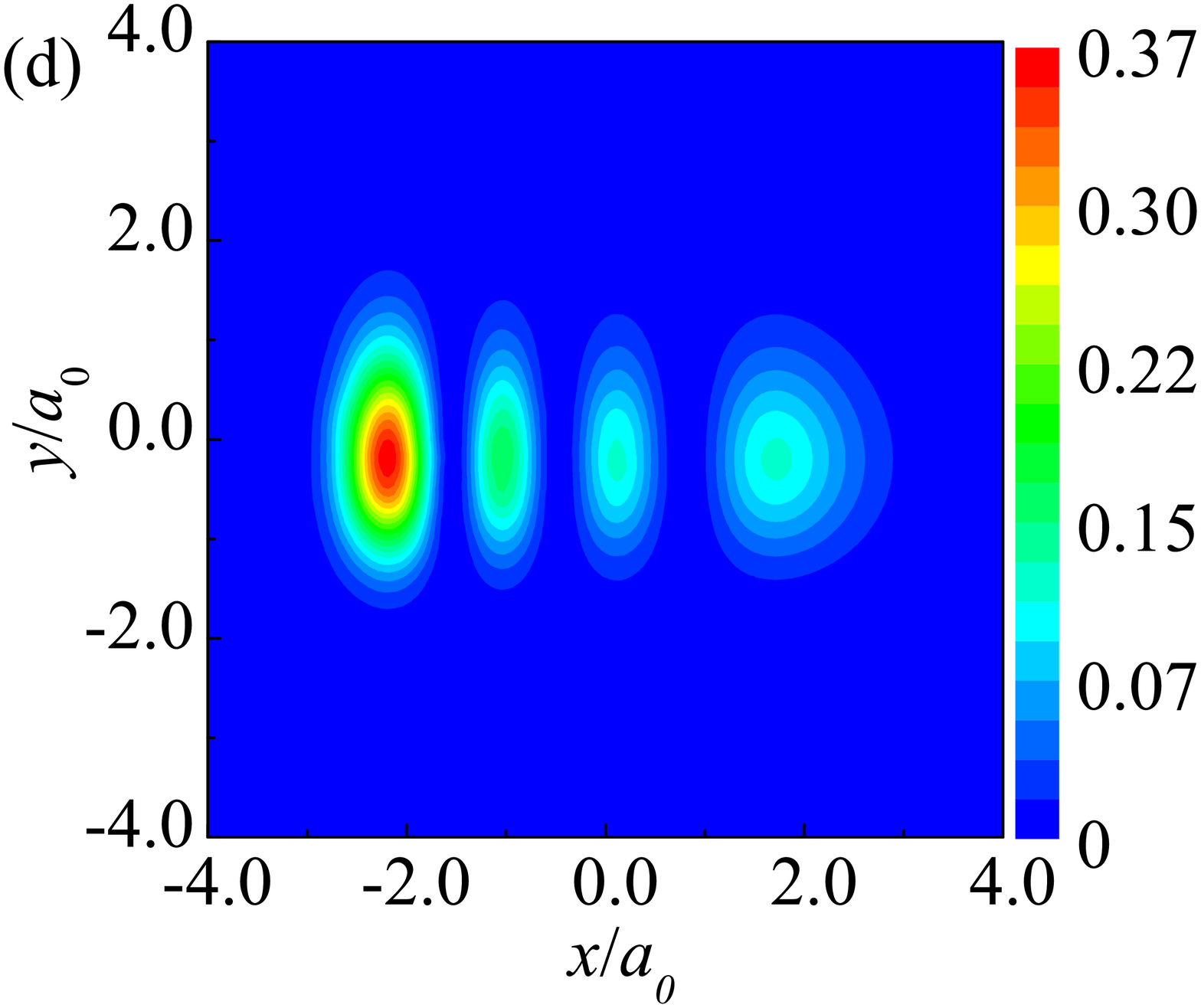}
  \end{minipage}
  \\
  \begin{minipage}[h]{0.45\linewidth}
   \includegraphics[width=\linewidth]{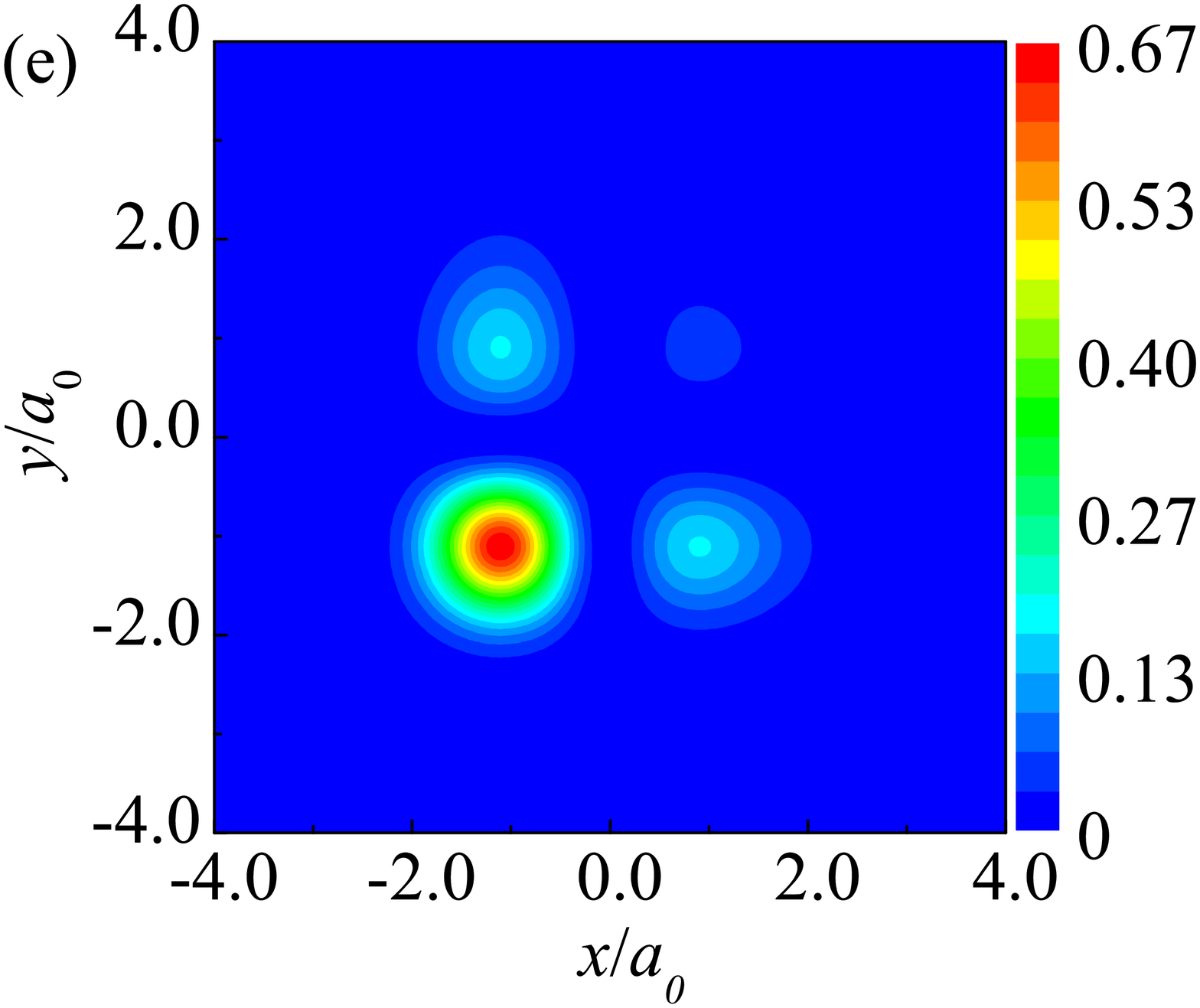}
  \end{minipage}
  \begin{minipage}[h]{0.45\linewidth}
   \includegraphics[width=\linewidth]{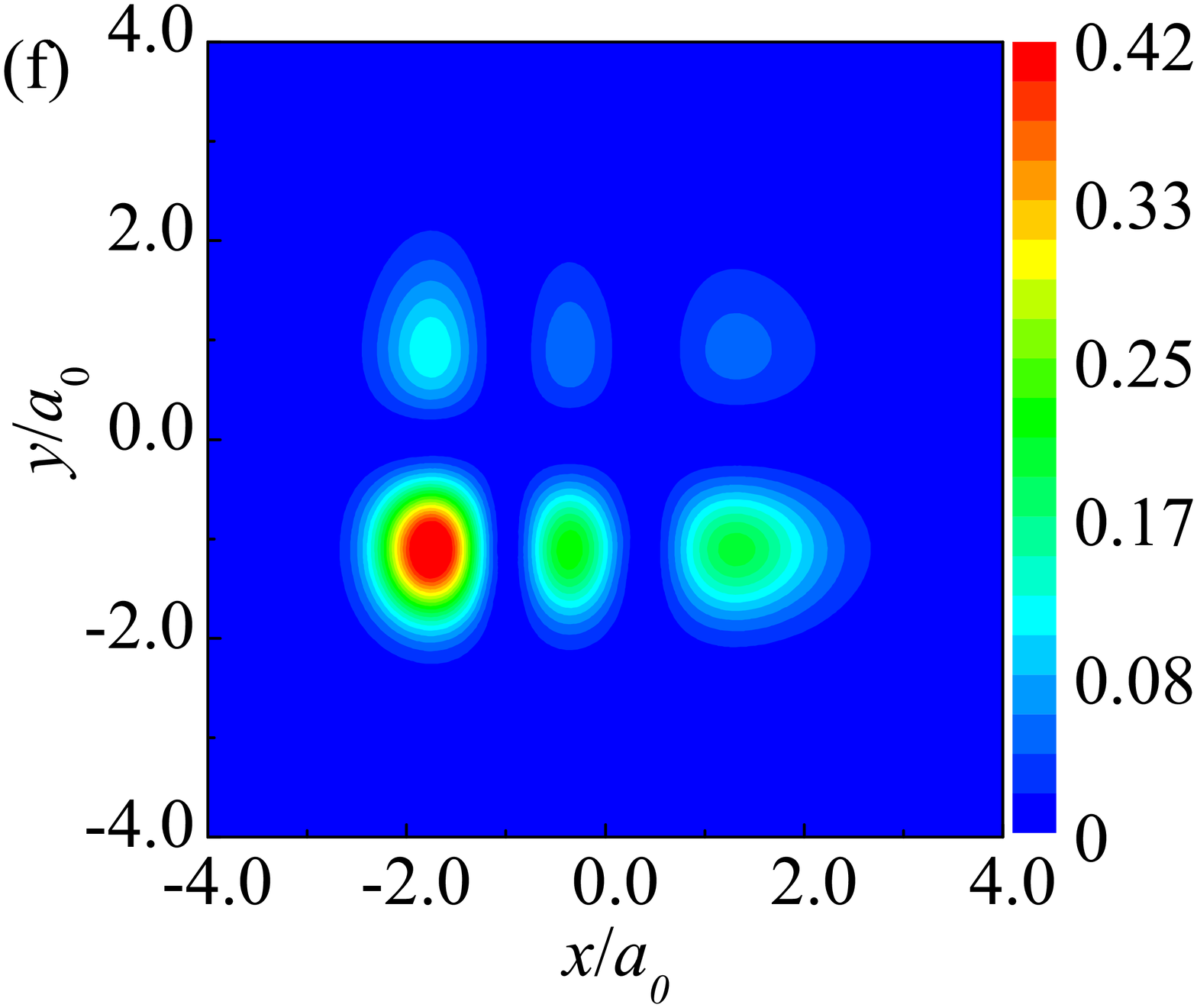}
  \end{minipage}
  \\
  \begin{minipage}[h]{0.45\linewidth}
   \includegraphics[width=\linewidth]{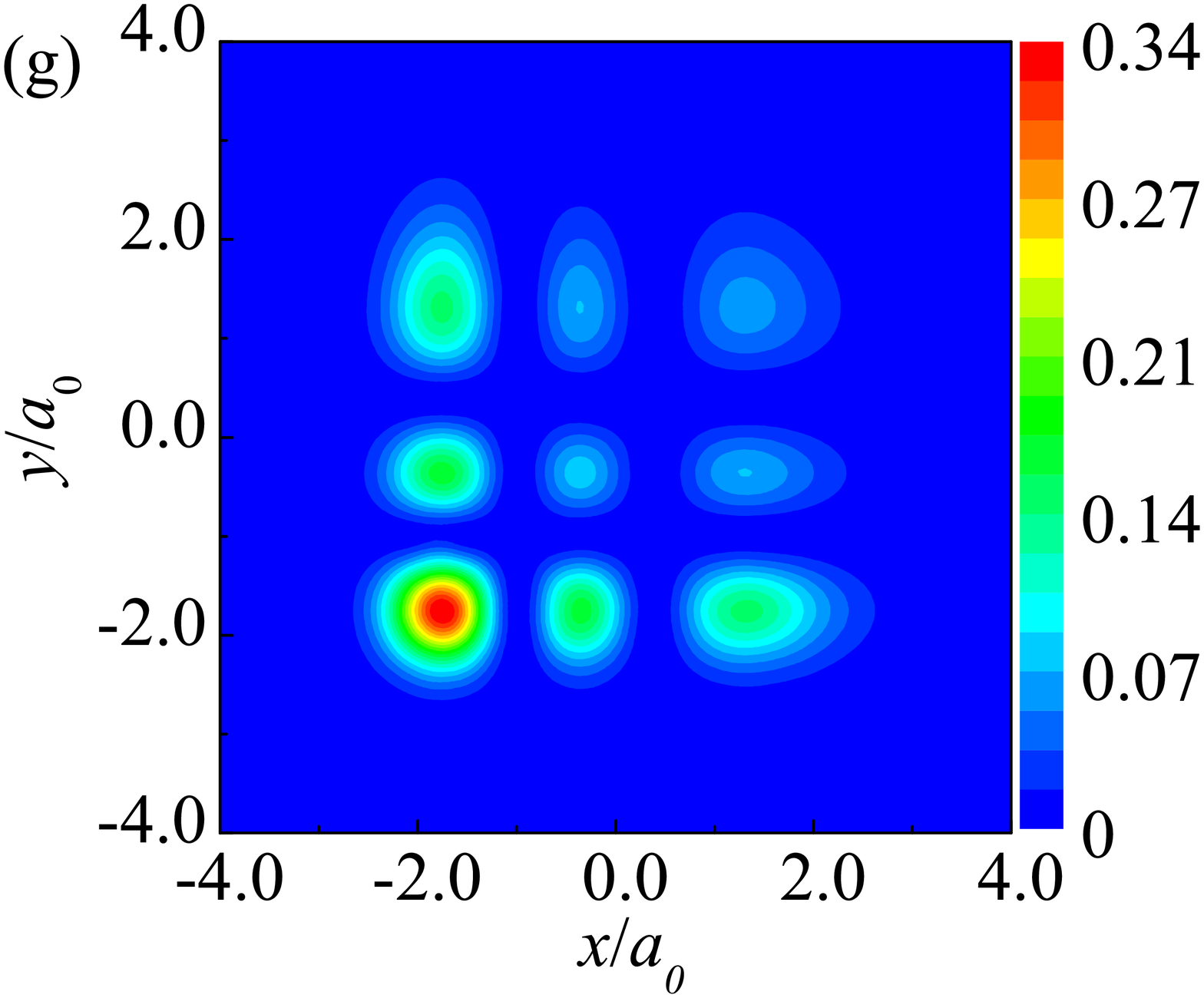}
  \end{minipage}
  \begin{minipage}[h]{0.45\linewidth}
   \includegraphics[width=\linewidth]{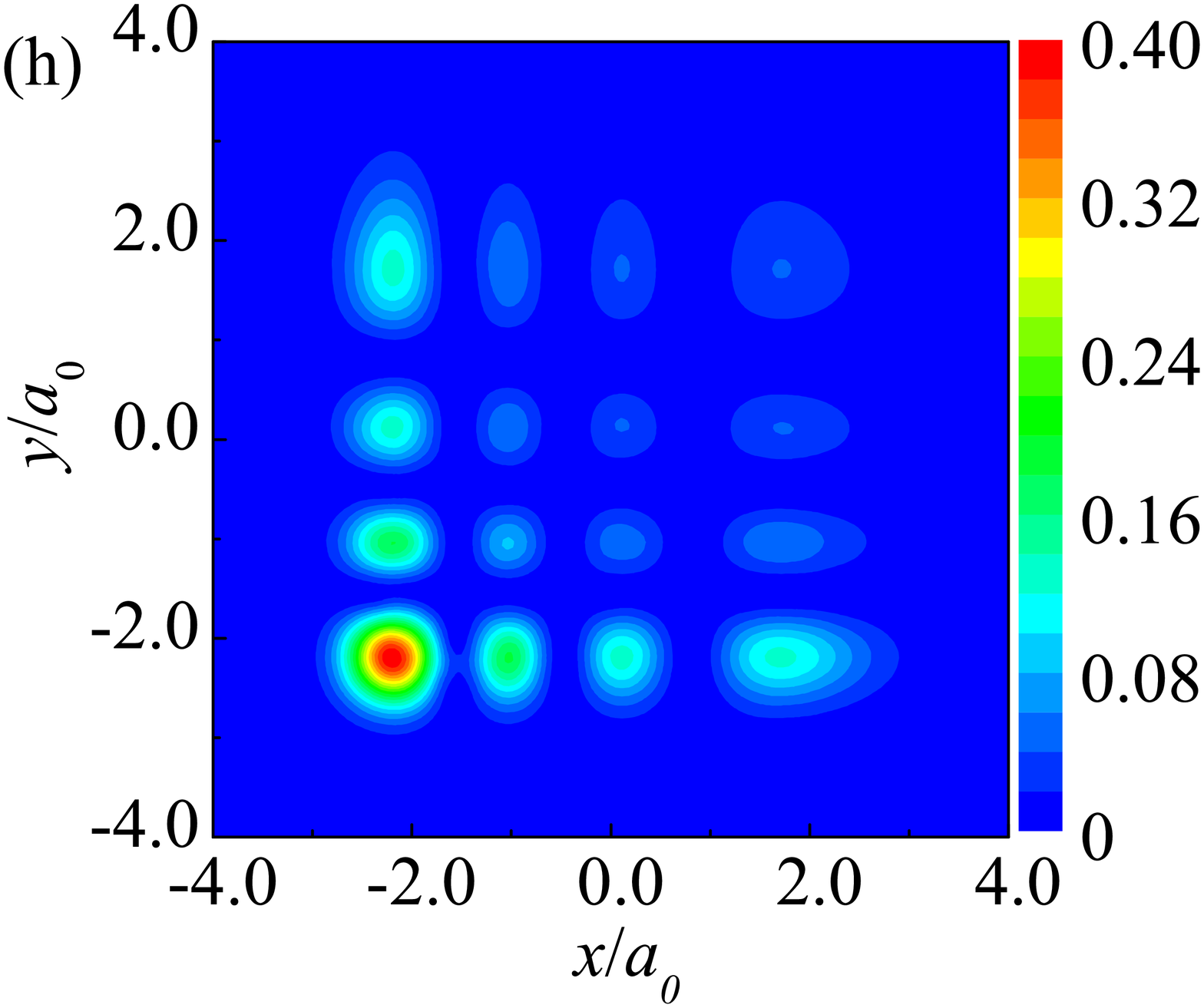}
  \end{minipage}
 \end{minipage}
 \caption{\label{fig:psi-osc-quant-2D-2}
  (Color online)
  Probability densities
  $\rho_{n_1, n_2}(x,y) = |\psi_{n_1} (x) \psi_{n_2} (y)|^2$ 
  (in units of $a_0^{-2}$)
  for $\gamma_q a_0 = 0.2$. 
  (a) $(n_1, n_2) = (0,0)$, (b) $(1,0)$, (c) $(2,0)$, (d) $(3,0)$,
  (e) $(1,1)$, (f) $(2,1)$, (g) $(2,2)$,  (h) $(3,3)$.
  Color scale ranges from blue (low probabilities) to red (high probabilities).
}
\end{figure}

Figure \ref{fig:psi-osc-quant-2D-2}
shows probability densities
$\rho_{n_1, n_2}(x,y) = |\psi_{n_1} (x) \psi_{n_2} (y)|^2$
for the two-dimensional case, where the asymmetric effect 
due to the PDM can be seen.
The optical analog of this two-dimensional oscillator 
is related to the propagation of Gaussian beams in non-homogeneous media
\cite{Yariv}.

The expectation values of 
$ \langle \hat{x} \rangle $,
$ \langle \hat{x}^2 \rangle $,
$ \langle \hat{p} \rangle $, and
$ \langle \hat{p}^2 \rangle $
are 
\begin{subequations}
 \label{eq:x-p-x^2-p^2-expec-values}
 \begin{equation}
  \label{eq:x-med-quantum}
  \langle \hat{x} \rangle = -\frac{\gamma_q \hbar}{m_0\omega} 
                            \left( n + \frac{1}{2} \right),
 \end{equation}
 \begin{equation}
  \label{eq:x^2-med-quantum}
  \langle \hat{x}^2 \rangle = \frac{\hbar}{m_0\omega} 
                              \left( n + \frac{1}{2} \right),
 \end{equation}
 \begin{equation}
  \label{eq:valor-expected-p}
  \langle \hat{p} \rangle = 0,
 \end{equation}
 \begin{eqnarray}
  \label{eq:valor-expected-p^2}
  \langle \hat{p}^2 \rangle =
 	m_0 \omega \hbar  \;
 	\frac{\left( n + \frac{1}{2} \right) 
              - \frac{\gamma_{q}^2 x_{0}^2}{2} (n^2 + n - 1)
             }
             {
              \left[ 
                    1 - \gamma_{q}^2 x_{0}^2\left( n + \frac{1}{2} \right) 
              \right]^2
              - (\gamma_{q}x_{0})^4
             }.
 \end{eqnarray}
\end{subequations}

We can clearly see that in the limit $\gamma_q x_0 \rightarrow 0$,
the usual cases
$\langle \hat{x} \rangle = 0 $ and
$\langle \hat{p}^2 \rangle = m_0 \omega \hbar \,(n + \frac{1}{2})$
are recovered.
Since $b/2d = \sqrt{1 - \gamma_q^2 a_{q,n}^2} $,
Eqs.~(\ref{eq:x-med-quantum}), 
(\ref{eq:x^2-med-quantum}) and
(\ref{eq:valor-expected-p^2})
can be rewritten as 
\begin{subequations}
\label{eq:moments-oscillator-quantum-amplitude}
\begin{equation}
\label{eq:x-oscillator-quantum-amplitude}
  \frac{\langle \hat{x} \rangle}{a_{q,n}} = 
	- \frac{1-\sqrt{1 - \gamma_q^2 a_{q,n}^2}}{\gamma_q a_{q,n}},
\end{equation}
\begin{equation}
\label{eq:x^2-oscillator-quantum-amplitude}
 \frac{\langle \hat{x}^2 \rangle}{a_{q,n}^2} 
	= \frac{1-\sqrt{1 - \gamma_q^2 a_{q,n}^2}}{\gamma_q^2 a_{q,n}^2},
\end{equation}
\begin{equation}
\label{eq:p^2-oscillator-quantum-amplitude}
	\langle \hat{p}^2 \rangle = \frac{m_0^2 \omega_0^2}{2}  
	\left(
	\frac{
	a_{q,n}^2  
	+ \frac{3}{4} \gamma_q^2 x_0^4
	}{
	\displaystyle
	1- \gamma_q^2 a_{q,n}^2 
	- \gamma_q^4 x_0^4
	} \right).
\end{equation}
\end{subequations}
According to the principle of correspondence,
in the limit of large quantum numbers (or $\hbar \rightarrow 0$), 
we have $E_n \rightarrow E$ and $a_{q,n} \rightarrow A$, and consequently 
Eqs.~(\ref{eq:moments-oscillator-quantum-amplitude})
coincide with 
Eqs.~(\ref{eq:x-p-x^2-p^2-medium}).
Similarly to the classic case, one obtains
$
 \langle \hat{T} \rangle
          = \sqrt{1 - \gamma_q^2 a_{q,n}^2} 
            \;
            \langle \hat{V} \rangle.
$

\begin{figure}[!htb]
 \centering
 \begin{minipage}[h]{0.52\linewidth}
  \includegraphics[width=\linewidth]{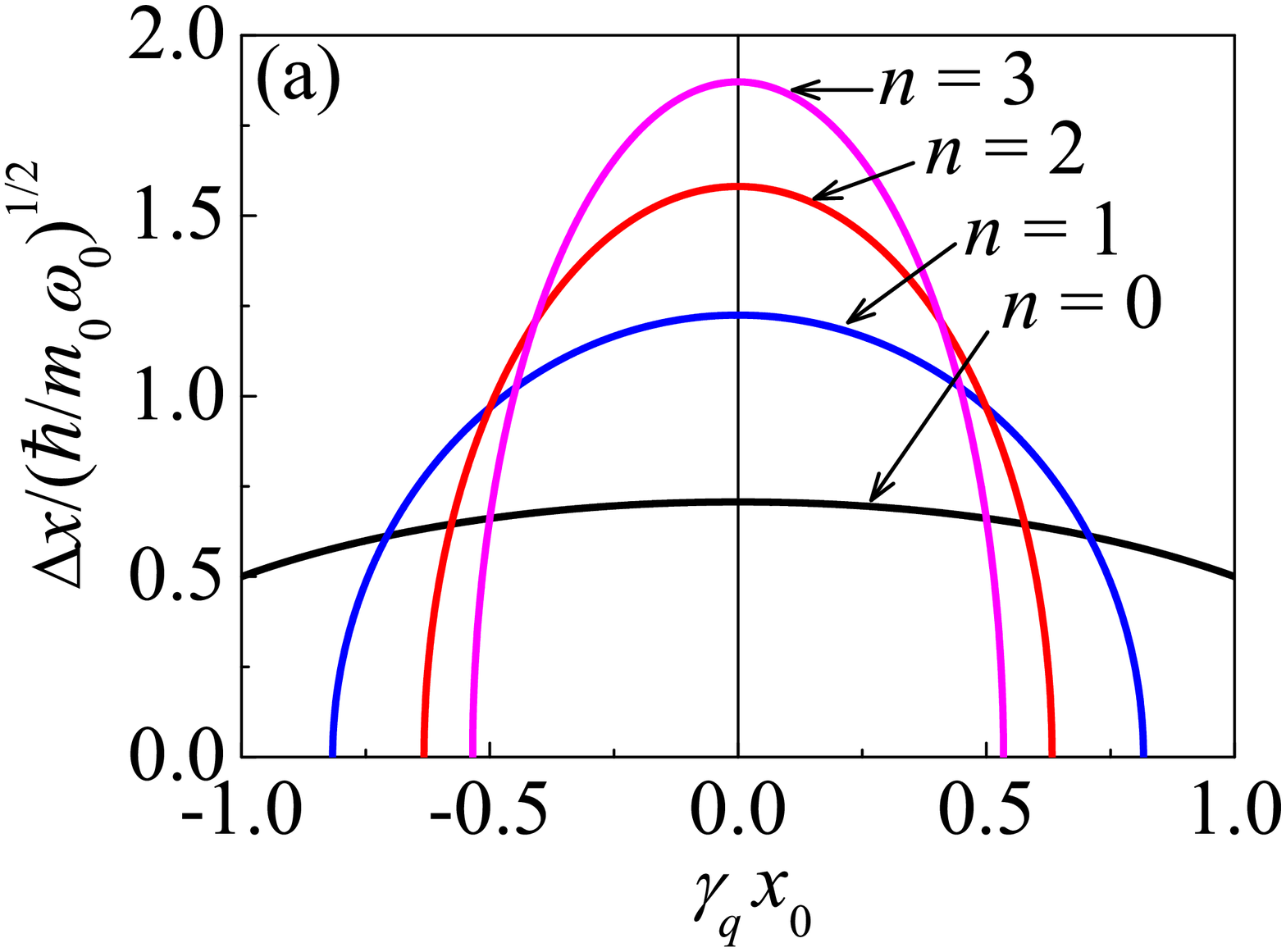}
 \end{minipage}\\
 \begin{minipage}[h]{0.52\linewidth}
  \includegraphics[width=\linewidth]{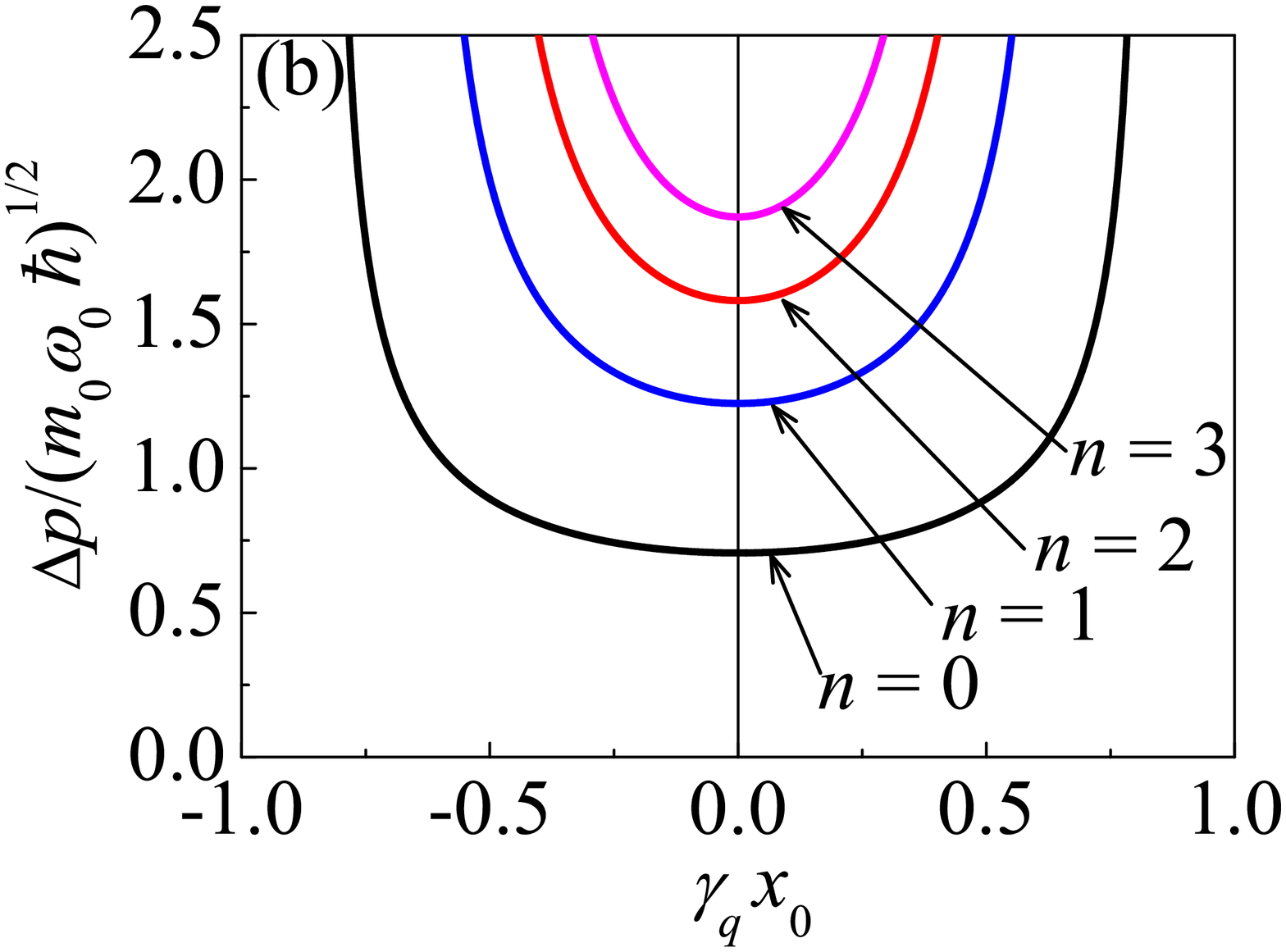}
 \end{minipage}\\
 \begin{minipage}[h]{0.52\linewidth}
  \includegraphics[width=\linewidth]{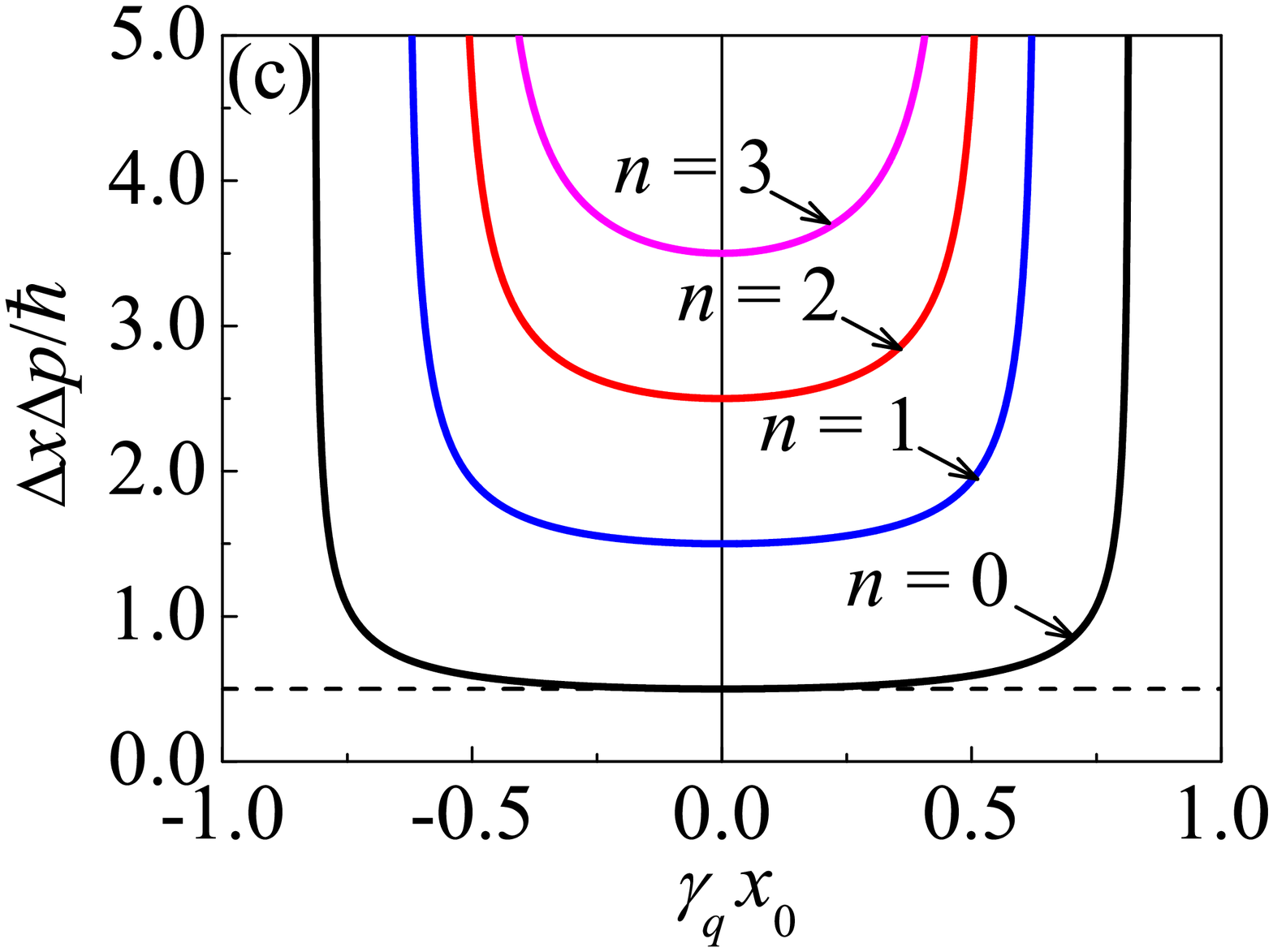}
 \end{minipage}
 \caption{\label{fig:uncertainty} 
  (Color online)
  Uncertainty of 
  (a) position $(\Delta x)$,
  (b) linear momentum $(\Delta p)$,
  (c) $\Delta x \Delta p$,
  for the first quantum numbers.
  The usual case $\Delta x \Delta p = (n+\frac{1}{2})\hbar$
  is recovered for $\gamma_q x_0 \rightarrow 0$.
 }
\end{figure}

Figure \ref{fig:uncertainty} shows the uncertainty relation for this oscillator.
$\Delta x \Delta p \geq  \hbar/2$, since that the operators 
$\hat{x}$ and $\hat{p}$ are canonically conjugate and Hermitian.

\section{Conclusions}

We revisit the problem of a particle with position-dependent mass 
introduced in \cite{costa-filho-2013}, where we use the Hermitian deformed 
linear momentum operator $\hat{p}_q$, 
instead of the originally proposed non-Hermitian operator.
The definition of a deformed space operator
$\hat{x}_q  = \xi \ln[\exp_q ( \hat{x}/\xi) ]$
(Eq.~(\ref{eq:operator-position-generalized}))
establishes the equivalence of the PDM 
particle in an ordinary phase space
to a constant mass particle in a deformed phase space.
Particularly, the PDM $m(x) = m_0/(1+\gamma_q x)^2$
(Eq.~(\ref{eq:m(x)}))
in an ordinary space submitted to a quadratic potential is transformed 
into a constant mass in a deformed space submitted to the Morse potential.
Since the operators $\hat{x}_q$ and $\hat{p}_q$ are Hermitian and
canonically conjugated, the classical and quantum formalisms
are coherently connected, and the uncertainty and correspondence principles
are consistently followed.
The introduction of deformed derivatives allows the equations of motion 
to be written with a formal similarity to the usual ones:
the quantum case uses the $q$-derivative, Eq.~(\ref{eq:q-derivative})
(see Eq.~(\ref{eq:deformed-schroedinger-equation})),
while the classical case uses the dual $q$-derivative,
Eq.~(\ref{eq:q-derivative-dual})
(see Eq.~(\ref{eq:second_newton_law_generalized})).
The specific PDM we have considered here, Eq.~(\ref{eq:m(x)}),
is associated with the deformed mathematical framework 
addressed in Sec.~\ref{sec:q-algebra}.
Investigations of the deformed oscillator through 
factorization methods in quantum mechanics, 
like supersymmetry 
\cite{plastino-1999,Amir-2016}
and generalized Heisenberg algebras
\cite{Curado-RegoMonteiro-2001,Curado-RegoMonteiro-Rodrigues-2013},
and how the operators that emerge from these formulations 
can be rewritten in terms of $q$-deformed functions and derivatives
are interesting additional developments.
The equivalence between the use of deformed variables
and deformed functions is already present in the works of Kaniadakis,
e.g., Eq. (50) of 
\cite{kaniadakis-physa-2001}
and Eq. (2.1) of 
\cite{kaniadakis-pre-2002}, 
though in a different context.
Other possible PDM functions may be connected to different 
deformations of space, and consequently to different derivatives,
and ultimately, deformed equations of motion.
Investigations on these possible unexpected connections are stimulating.

\begin{acknowledgments}
We thank M.\ A.\ Rego-Monteiro and F.\ F.\ dos Santos 
for fruitful discussions.
This work was partially supported by FAPESB (Brazilian agency), 
through the program PRONEX.
\end{acknowledgments}



\end{document}